\documentclass[aps, reprint, superscriptaddress, amsmath, amssymb]{revtex4-1}

\usepackage{CJK}
\usepackage{graphicx}
\usepackage{mathtools}
\usepackage{multirow}
\usepackage{braket}

\usepackage[usenames, dvipsnames]{color}
\usepackage{hyperref}
\hypersetup{colorlinks, citecolor=NavyBlue, linkcolor=ForestGreen, urlcolor=Fuchsia}
\usepackage{soul}

\begin{document}

\title{Collective resonances of atomic xenon from the linear to the nonlinear regime}

\begin{CJK*}{UTF8}{}

\author{Yi-Jen Chen \CJKfamily{bsmi}(陳怡蓁)}
\email{yi-jen.chen@desy.de}
\affiliation{Center for Free-Electron Laser Science, DESY, Notkestrasse 85, 22607 Hamburg, Germany}
\affiliation{Department of Physics, University of Hamburg, Jungiusstrasse 9, 20355 Hamburg, Germany}

\author{Stefan Pabst}
\affiliation{Center for Free-Electron Laser Science, DESY, Notkestrasse 85, 22607 Hamburg, Germany}
\affiliation{ITAMP, Harvard-Smithsonian Center for Astrophysics, 60 Garden Street, Cambridge, MA 02138, U.S.A.}

\author{Robin Santra}
\email{robin.santra@desy.de}
\affiliation{Center for Free-Electron Laser Science, DESY, Notkestrasse 85, 22607 Hamburg, Germany}
\affiliation{Department of Physics, University of Hamburg, Jungiusstrasse 9, 20355 Hamburg, Germany}

\date{\today}

\begin{abstract}
XUV nonlinear spectroscopy has recently discovered that there is more than one collective dipole resonance state in the energy range of the giant dipole resonance (GDR) of atomic Xe. This resonance-state substructure, hidden in the linear regime, raises imminent questions regarding our understanding of the collective electronic behavior of Xe, which has been largely founded on linear spectroscopic studies. Here, we approach the collective response of Xe from a new perspective: we study directly the resonance eigenstates, and then analyze their spectroscopic manifestations. We find that linear spectroscopy captures only partial information on the resonance substructure as a result of quantum interferences. Moreover, we show that the resonance state dominating the GDR in linear spectroscopy has no adiabatic connection to the resonance state governing the corresponding cross section when multielectron interactions are neglected. Going beyond the dipole-allowed correlated electronic structure, we predict the existence of collective multipole resonances of Xe. Unlike any known collective feature in atoms, these resonances live exceptionally long (more than 100 attoseconds), thus providing a new playground for studying the collective nonlinear response of Xe using advanced light sources.

\null \noindent Keywords: inner-shell photoionization, multiphoton ionization, electronic correlations, ab initio calculations, free-electron lasers, attosecond dynamics
\end{abstract}


\maketitle

\end{CJK*}

\section{\label{Sec:Intro}Introduction}
Although the one-particle approximation can well describe a multielectron atom in several aspects, it fails conspicuously in some cases to capture the many-body nature of an atom. Indeed, as early as 1933, Bloch proposed the existence of plasma-like collective excitations within an atom \cite{f.bloch33a}. It is well known nowadays that such collective excitations can take place during XUV one-photon ionization of heavy atoms, best showcased by the giant dipole resonance (GDR) of Xe \cite{m.amusia90b, m.amusia00a, c.brechignac94a, j.connerade87a}. Upon photoabsorption, the $f$-wave photoelectron promoted from the inner $4d$ subshell is temporarily trapped close to the ion \cite{a.starace82a}, which then sets off strong particle-hole interactions. This cooperative electronic motion is typically reflected in calculations of the one-photon ionization cross section: a vast amount of theoretical investigations has shown the importance of correlations for the quantitative agreement between theory \cite{j.cooper64a, m.amusia67a, w.brandt67a, a.starace70a, g.wendin73c, a.zangwill80a} and experiment \cite{d.ederer64a, a.lukirskii64a}.

Even today, the GDR remains captivating, since its collective character lies at the heart of the nonlinear response of Xe to various new light sources. Not only does it lead to a striking enhancement in the high-harmonic generation (HHG) spectrum of Xe driven by intense NIR lasers \cite{a.shiner11a, s.pabst13b}, but it also creates an unusual charge-state distribution of Xe irradiated by XUV free-electron lasers (FELs) \cite{m.richter09a, n.gerken14a}.

More recently, an experiment has been performed at the XUV-FEL FLASH on two-photon above-threshold ionization (ATI) of Xe, taking the GDR as an intermediate step \cite{t.mazza15a}. The photoemission yields were recorded as a function of the FEL intensity at two selected photon energies. A many-body theory well reproduced the measurements, and was in turn used for an in-depth study of the photoionization cross sections over a wide energy range \cite{t.mazza15a}. The two-photon cross section, surprisingly, unveiled a prominent knee-type structure with two kinks, which is in sharp contrast to the single smooth hump observed in the one-photon cross section. This result provides strong evidence that there is more than one dipole-allowed, collective resonance state in the energy range of the GDR---that there are in fact two sub-resonances \cite{g.wendin73c, y.chen15a}. A question immediately surfaces: why is this resonance-state substructure, a key indicator of electronic correlations \cite{t.mazza15a, y.chen15a}, not detected by linear spectroscopy, a central tool for our understanding of the collective behavior of Xe over the past half-century \cite{m.amusia90b, m.amusia00a, c.brechignac94a, j.connerade87a, a.starace82a, j.cooper64a, m.amusia67a, w.brandt67a, a.starace70a, g.wendin73c, a.zangwill80a, d.ederer64a, a.lukirskii64a}?

In this paper, we answer this question by resorting to the \textit{fundamental electronic structure} using the concept of \textit{resonance eigenstates}. We show that only one of the two sub-resonances can be resolved by linear spectroscopy, while the other is nearly invisible as a consequence of quantum interferences. Because the linear response of Xe only reveals \textit{partial} information on the correlated electronic structure, there is no contradiction between the resonance substructure and the structureless hump in the one-photon cross section. In addition, we demonstrate how many-body effects result in the emergence of the two collective sub-resonances from the one-particle resonances. It turns out that the birthing process of the correlated electronic structure deviates substantially from what would be expected from conventional linear response theory. Furthermore, we predict the existence of hitherto unknown collective multipole resonance states of the $4d^{-1} \epsilon f$ type, the first unambiguous case of atomic plasmons on account of their comparatively long lifetimes. New prospects are now in sight for probing the multielectron dynamics of Xe with XUV nonlinear spectroscopy or attosecond metrology.

Before proceeding further, it is necessary to clarify what we mean by a resonance. In the theory of resonances, a resonance can be strictly identified as a discrete eigenstate of the Hamiltonian with a complex eigenenergy \cite{n.moiseyev11a, n.moiseyev98b, y.ho83a, w.reinhardt82a, r.santra02a}. We therefore distinguish between a resonance in terms of the electronic structure and a resonance-like feature in the photoabsorption spectra; the latter is simply a possible manifestation of the former. Nonetheless, as a matter of convention, the acronym ``GDR'' is still used here to refer to the single smooth maximum in the one-photon ionization cross section of Xe.

This work extends the literature on collective electronic excitations in and beyond atomic systems in two important ways. First, we provide a unique \textit{bottom-up} approach to interpret the spectroscopic data of collective excitations. We study the collective resonance states \textit{explicitly}, which enables an analysis of the cross sections on a state-by-state basis. Conversely, prior research adopts a \textit{top-down} approach, where the resonance states are inferred from spectroscopic features \cite{m.amusia90b, m.amusia00a, c.brechignac94a, j.connerade87a, a.starace82a}. Standard linear response theory applying advanced many-body techniques has made remarkable achievements in the quantitative description of the one-photon absorption data of the GDR \cite{m.amusia67a, w.brandt67a, a.starace70a, g.wendin73c, a.zangwill80a, z.altun88a}. Nevertheless, the fact that spectroscopic signals are knotty outcomes of quantum interferences among overlapping resonance and continuum states undermines their ability to reveal not only the full structural information but also the role of correlations in shaping the electronic structure.

Second, our study is, to the best of our knowledge, the first successful attempt to find collective multipole excitations in a realistic many-electron system. As far as experiment is concerned, the collective resonances of Xe under consideration are still confined to the dipole mode(s) \cite{d.ederer64a, a.lukirskii64a, a.shiner11a, m.richter09a, n.gerken14a, t.mazza15a}. On the theory side, Ref.~\cite{n.cherepkov01a} examines the quadrupole transition matrix elements related to the $np^{-1} \epsilon f$ resonances of Xe in a many-body framework. Yet, as the number of $np$ electrons is much less than that of $4d$, the resonances there are basically of one-particle origin \cite{n.cherepkov01a}. Using a mean-field model, Ref.~\cite {l.pi14a} peruses the generalized cross sections connected to the $4d^{-1} \epsilon f$ octupole resonance of Xe. According to the dipole case, it is no surprise that adding electronic correlations will shift the parameters of the noninteracting multipole resonances. What is unanticipated, however, is the direction in which these modifications unfold: as will be shown below, multielectron effects actually sharpen the widths of the $4d^{-1} \epsilon f$ multipole resonances of Xe insomuch that they end up being at least 20 eV narrower than their dipole counterparts.

The remainder of this paper is organized as follows. Section \ref{Sec:Theory} lays out the theoretical methodology. Section \ref{Subsec:Ses} characterizes the fundamental electronic structure of interest, namely the $4d^{-1} \epsilon f$ resonance states of Xe. Section \ref{Subsec:SesEvo} illustrates how many-electron interactions trigger the rearrangement of the electronic structure and give birth to the collective resonances. Section \ref{Subsec:OnePhotonCx} shows how to dissect the one-photon cross section in a bottom-up, state-resolved manner and explains why quantum interferences block out the visibility of one sub-resonance in the vicinity of the GDR. Section \ref{Subsec:ThreePhotonCx} presents a three-photon--two-color scheme that is designed specifically for probing the collective octupole resonance uncovered in this work. Section \ref{Sec:Conclusion} concludes our paper. Further information on the numerical details and on the properties of the one-particle resonances can be found in Appendix \ref{Sec:NumDtl} and Appendix \ref{Sec:OneBodyRes}, respectively. Atomic units (a.u.) are used throughout the article ($|e| = m_e = \hbar = 4 \pi \epsilon_0=1$) unless otherwise stated.

\section{\label{Sec:Theory}Theory}
Here, we investigate the \textit{ab initio} electronic structure in the $4d$ continuum of Xe by diagonalizing the $N$-electron Hamiltonian subjected to smooth exterior complex scaling (SES) \cite{n.moiseyev98a, h.karlsson98a, c.buth07b, s.pabst16b} within the wave-function--based configuration interaction singles (CIS) many-body theory \cite{n.rohringer06a, l.greenman10a, s.pabst12b, s.pabst13a, s.pabst17a}. This combination has already been described in Ref.~\cite{y.chen15a}, and has successfully predicted the dipole-accessible resonance substructure \cite{y.chen15a} in consistence with the knee-type structure in the two-photon ATI cross section \cite{t.mazza15a}.

In CIS, we represent the nonrelativistic $N$-body Hamiltonian including the exact two-electron Coulomb interactions in the CIS configuration space $\mathcal{V}_{\text{CIS}}$, which comprises the Hartree-Fock (HF) ground state $\ket{\Phi^{\text{HF}}_0}$ and its singly excited one-particle--one-hole (1p-1h) configurations $\ket{\Phi^\mathbf{a}_\mathbf{i}}$ \cite{a.szabo96a}. As the wave function ansatz is a summation over Slater determinants, CIS is able to capture essential correlation physics beyond the mean-field level \cite{n.rohringer06a, l.greenman10a, s.pabst12b, s.pabst13a, s.pabst17a}. Applications of CIS to processes involving the GDR can be found in Refs.~\cite{s.pabst13b, t.mazza15a, y.chen15a, d.facciala16a, d.krebs14a, s.pabst15a}. In particular, the quantitative capability of CIS can be seen in Refs.~\cite{t.mazza15a, d.facciala16a, d.krebs14a}.

Resonances belong to a special type of electronic structure. Due to their asymptotic divergent behavior, they are not in the Hilbert space of a Hermitian Hamiltonian \cite{n.moiseyev98b, r.santra02a}. To circumvent this problem, SES \cite{n.moiseyev98a, h.karlsson98a, c.buth07b, s.pabst16b}, a variant of the well-established complex scaling theory \cite{n.moiseyev11a, n.moiseyev98b, y.ho83a, w.reinhardt82a}, is employed to rigorously transform a resonance state into one single square integrable, bound-state--like eigenfunction of the scaled non-Hermitian Hamiltonian. Note that the symmetric inner product $\left( \cdot \right| , \left| \cdot \right)$ instead of the Hermitian one $\left< \cdot \right| , \left| \cdot \right>$ must be used to assure orthogonality among the eigenvectors of the scaled Hamiltonian \cite{n.moiseyev11a, n.moiseyev98b}. Complex scaling is usually used to address resonance phenomena in few-electron systems \cite{n.moiseyev11a, n.moiseyev98b, y.ho83a, w.reinhardt82a, a.scrinzi00a, d.telnov02a, x.bian11a}, but had not been used before for collective excitations in a multielectron atom as complex as Xe.

Exploiting the conservation of the total spin and total magnetic quantum numbers, the hole index $\mathbf{i} = nl_{\pm m}$ specifies one ionization channel \cite{n.rohringer06a}, and $\mathcal{V}_{\text{CIS}}$ accommodates only spin-singlet 1p-1h configurations reachable by single or multiple dipole excitations of $\ket{\Phi^{\text{HF}}_0}$ in linearly polarized light fields \cite{s.pabst12b}. The computations are carried out using our \textsc{XCID} package \cite{xcid}. For numerical parameters please refer to Appendix \ref{Sec:NumDtl}.

To systematically assess the many-body effects, we often compare the results of two scenarios: the full CIS model and a reduced intrachannel model. The genuine two-electron correlations within CIS are fully encapsulated by the interchannel-coupling Coulomb matrix elements $( \Phi^\mathbf{b}_\mathbf{j} | \hat{H}_{\text{e-e}} | \Phi^\mathbf{a}_\mathbf{i} )$ with $\mathbf{i} \neq \mathbf{j}$ \cite{a.starace82a, s.pabst12b}. It is this type of interactions that can simultaneously change the state of the photoelectron ($\mathbf{a} \rightarrow \mathbf{b}$) and that of the cation ($\mathbf{i} \rightarrow \mathbf{j}$), enabling the formation of an entangled particle-hole pair \cite{s.pabst17a}. In the intrachannel model, all the interchannel terms are set to zero, and only the intrachannel terms $( \Phi^\mathbf{b}_\mathbf{i} | \hat{H}_{\text{e-e}} | \Phi^\mathbf{a}_\mathbf{i} )$ are considered \cite{a.starace82a, s.pabst12b}. The intrachannel Hamiltonian effectively acts as a one-particle nonlocal potential \cite{s.pabst17a}.

\section{\label{Sec:Results}Results}

\subsection{\label{Subsec:Ses}Characterization of the resonance eigenstates}
\begin{figure}
\includegraphics[angle=270]{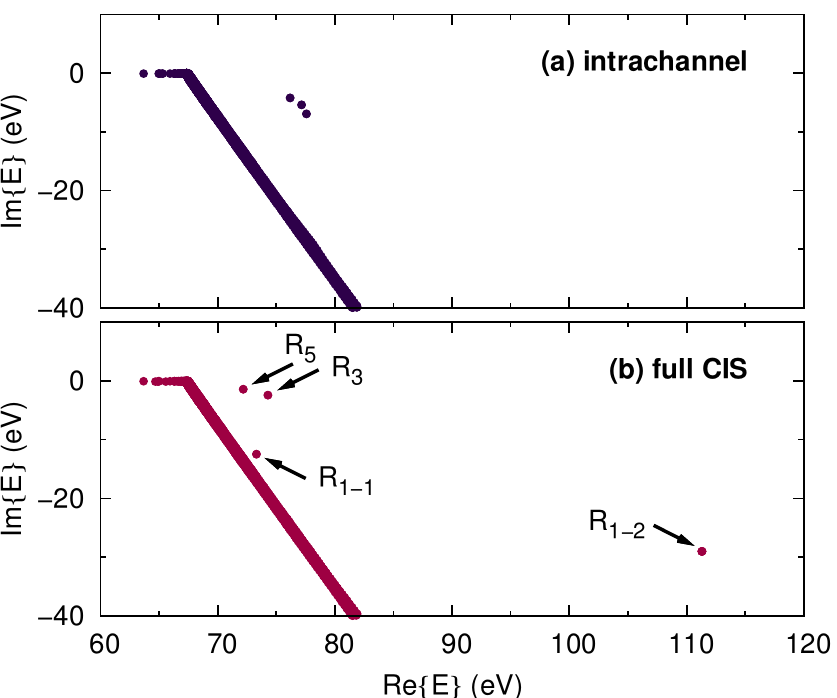}
\caption{\label{Fig:Ses}Complex energy spectra near the $4d$ threshold of Xe in (a) the intrachannel and (b) the full CIS models. Each filled circle symbolizes an eigenstate of the scaled $N$-electron Hamiltonian. Horizontal and vertical axes represent the real and imaginary parts of the energy eigenvalue, respectively.}
\end{figure}

Figure \ref{Fig:Ses}(a) presents the spectrum of the energy eigenvalues for the complex-scaled, non-Hermitian intrachannel Hamiltonian in close proximity to the $4d$ ionization threshold at 67.5 eV \cite{a.thompson09a}. As stated by the Balslev-Combes theorem \cite{n.moiseyev11a, n.moiseyev98b}, the bound states remain on the real energy axis, the continuum is rotated clockwise by twice the scaling angle (an SES parameter \cite{y.chen15a}), and the resonances are exposed poles above the rotated continuum. The eigenenergy of each pole is the Siegert energy \cite{n.moiseyev98b, r.santra02a}: $E = \Xi - i \Gamma / 2$, with $\Xi$ the excitation energy and $\Gamma$ the inverse lifetime for the quasibound electron to escape to infinity. A group of three practically degenerate $4d^{-1}_{\pm m} \epsilon f_{\pm m}$ uncorrelated resonances \cite{a.starace82a, j.cooper64a} can be seen, one for each $4d_{\pm m}$ channel \cite{y.chen15a}. The tiny energy splitting hints at a slight dependency of the one-particle potential on the hole alignment.

Figure \ref{Fig:Ses}(b) depicts the energy spectrum for the full CIS model, where four resonances are visible. Clearly, the resonance substructure critically hinges on the two-body Coulomb interactions. The Siegert energies of the exposed resonances are detailed in Table \ref{Tab:SesEngy}. Each resonance state here has a definite total orbital angular momentum quantum number $L$. $R_{1-1}$ and $R_{1-2}$ [$4d^{-1} \epsilon f (^1P)$] are the two dipole sub-resonances in the spectral range of the GDR \cite{t.mazza15a, g.wendin73c, y.chen15a}. $R_{3}$ [$4d^{-1} \epsilon f (^1F)$] and $R_{5}$ [$4d^{-1} \epsilon f (^1H)$] were so far unknown and can be accessed only via multiphoton absorption. The $^1D$ and $^1G$ resonances are absent owing to the restrictions imposed on $\mathcal{V}_{\text{CIS}}$. Assuming electric dipole transitions for linearly polarized radiation \cite{f.faisal87a}, a closed-shell ground state is allowed to go to an odd-parity excited state with only an odd $L$. The hole population in each resonance state \cite{l.greenman10a} is primarily distributed among different $4d^{-1}_{\pm m}$, with small admixtures of $5s^{-1}_{0}$ and $5p^{-1}_{\pm m}$ from the outer shell \cite{m.amusia00a, a.starace82a, y.chen15a}. As a result of channel mixing, the resonance wave function must be written as a coherent superposition of various 1p-1h configurations and thereby represents a collective excitation \cite{b.povh95a}.

\begin{table}
\caption{\label{Tab:SesEngy}Siegert energies of the collective resonance states in the full CIS model. The energy values have an error bar of 0.5 eV, which is estimated by varying the SES parameters over a sensible range. For details please see Appendix \ref{Sec:NumDtl}.}
\begin{ruledtabular}
\begin{tabular}{lcccc}
Label & Configuration & $\Xi$ (eV) & $\Gamma$ (eV) & $\Gamma^{-1}$ (as)\\
\hline
$R_{1-1}$ & $4d^{-1} \epsilon f (^1P)$ & $73.3$ & $24.9$ & $26.4$\\
$R_{1-2}$ & $4d^{-1} \epsilon f (^1P)$ & $111.3$ & $58.0$ & $11.4$\\
$R_{3}$ & $4d^{-1} \epsilon f (^1F)$ & $74.3$ & $4.9$ & $135.4$\\
$R_{5}$ & $4d^{-1} \epsilon f (^1H)$ & $72.2$ & $2.8$ & $237.4$\\
\end{tabular}
\end{ruledtabular}
\end{table}

\subsection{\label{Subsec:SesEvo}Many-body effects on the emergence of the correlated electronic structure}
\begin{figure}
\includegraphics[angle=270]{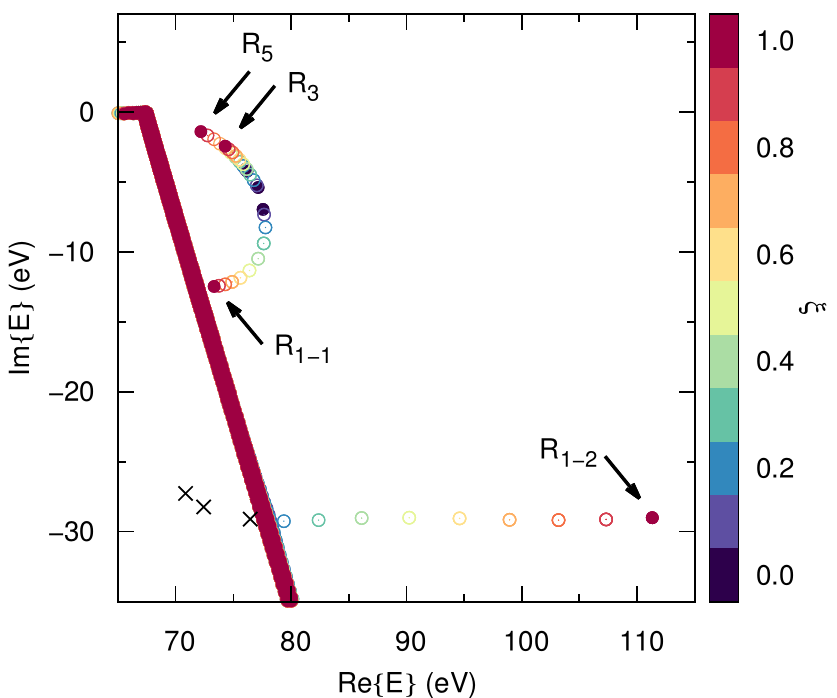}
\caption{\label{Fig:SesEvo}Evolution of the complex energy spectrum from the intrachannel to the full CIS model. The interchannel-coupling strength $\xi$ is represented by the false color. Locations of the unexposed intrachannel resonances are indicated by black crosses.}
\end{figure}

To elucidate the role of correlations in shaping the one-particle into the many-body resonance substructure, Fig.~\ref{Fig:SesEvo} illustrates how an eigenstate in the intrachannel model evolves into one in the full CIS model upon adiabatic switching of multielectron interactions. Briefly, we diagonalize the Hamiltonian gradually varying the strength of the interchannel-coupling terms $\xi ( \Phi^\mathbf{b}_\mathbf{j} | \hat{H}_{\text{e-e}} | \Phi^\mathbf{a}_\mathbf{i} )$, $\xi \in [0,1]$ ($\mathbf{i} \neq \mathbf{j}$). The intrachannel model is equivalent to the case where $\xi = 0$, and the full model to the case where $\xi = 1$.

First, we focus on the upper-left corner of Fig.~\ref{Fig:SesEvo}. As $\xi$ increases, the three intrachannel resonances turn into $R_{1-1}$, $R_{3}$, and $R_{5}$ in the full model. At first sight, one might picture this as multiplet splitting in the subspace spanned by the three one-particle resonances. However, the formation of $R_{1-1}$, $R_{3}$, and $R_{5}$ is far more complicated than that and requires configuration mixing among the intrachannel resonance and continuum states. A simple way to see this is that the eigenvalues of the reduced $3 \times 3$ Hamiltonian with off-diagonal interchannel couplings must analytically add up to a constant for an arbitrary $\xi$, whereas the sum of the $\Xi$'s for these three poles is obviously not conserved.

Interestingly, multielectron effects are particularly strong on the widths of the members of this resonance group, resulting in anomalous segregation (at least 20 eV in terms of $\Gamma$) of the broad dipole mode and the two narrow multipole modes. To date, no sustained collective excitation in atoms, or ``atomic plasmon'', has been found---it always carries charge density oscillations damping out more or less within a period \cite{m.amusia00a, c.brechignac94a}. As $R_{3}$ and $R_{5}$ are relatively long-lived (with $\Xi / \Gamma > 15$), they can be justly called ``atomic multipole plasmons'' and are expected to give rise to distinctive signatures in XUV nonlinear spectroscopy \cite{j.feldhaus10a, e.allaria10a, t.sekikawa04a} or attosecond pump-probe experiments \cite{f.krausz09a}.

Next, let us look at the lower part of Fig.~\ref{Fig:SesEvo}. When turning on couplings, the very broad resonance $R_{1-2}$ emerges from the continuum, retains its width, and quickly becomes fairly isolated in the energy plane. The emergence of $R_{1-2}$ does not mean that correlations create an additional resonance pole. Instead, it signals that the intrachannel potential can support another group of $4d^{-1}_{\pm m} \epsilon f_{\pm m}$ uncorrelated resonances with $\operatorname{Re}\{E\} \approx 70 \ \text{eV}$, $\operatorname{Im}\{E\} \approx -30 \ \text{eV}$ (see the black crosses in Fig.~\ref{Fig:SesEvo}) that is not exposed by the scaling angle chosen for the figures. With the angle used, these unexposed resonances are embedded in the intrachannel continuum close to $ \operatorname{Im}\{E\} = -30 \ \text{eV}$. The reason for keeping them unexposed is given in Appendix \ref{Sec:NumDtl}.

The two dipole sub-resonances $R_{1-1}$ and $R_{1-2}$ in the GDR region can be adiabatically traced back to two separate groups of one-particle resonances, whose existence was not known before. Note that it is rather unusual that a potential supports two resonance groups with distinct widths but similar excitation energies \cite{n.moiseyev98b, r.santra02a}. Ergo, the question naturally follows as to which properties of the one-particle potential are necessary for the presence of such a double-pole structure. We find that, while the Siegert energies of the exposed intrachannel resonances can be reasonably reproduced using only the direct part of the ionic potential, the appearance of the unexposed ones requires the nonlocal exchange part, which enters the intrachannel Coulomb matrix elements $( \Phi^\mathbf{b}_\mathbf{i} | \hat{H}_{\text{e-e}} | \Phi^\mathbf{a}_\mathbf{i} )$ to leading order through a dipolar term \cite{l.greenman10a, s.pabst12b}. Because of this nonlocality, the origin of the double-pole structure cannot be explained by the standard notion of shape resonances and local potential barriers \cite{a.starace82a} and thus remains nontrivial. For more information please see Appendix \ref{Sec:OneBodyRes}.

\subsection{\label{Subsec:OnePhotonCx}State-by-state analysis of the one-photon absorption cross section}
\begin{figure*}
\includegraphics[angle=270, width=6.7in]{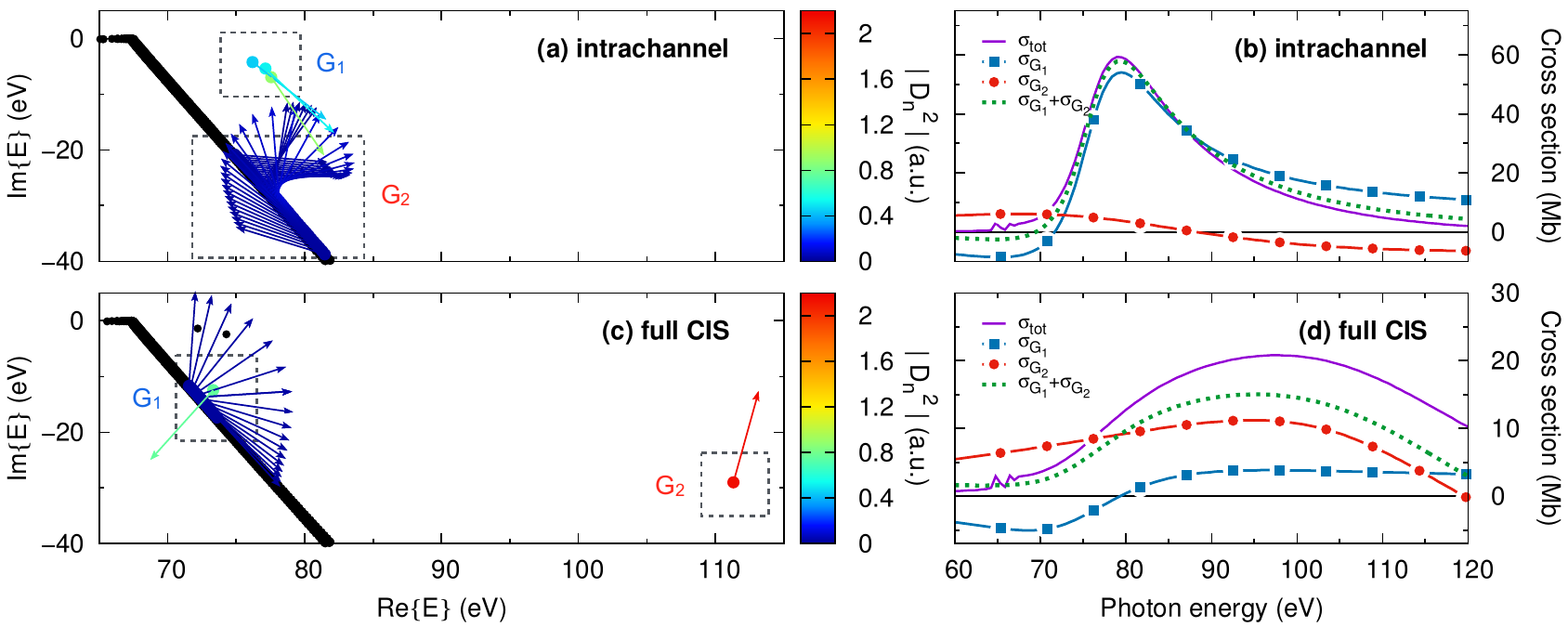}
\caption{\label{Fig:OnePhotonCx}Left panels: Distributions of ${D_n}^2$ in (a) the intrachannel and (c) the full CIS models. Each state with $|{D_n}^2| > 0.016$ is represented by a filled circle and a vector of constant length. The amplitude $|{D_n}^2|$ is indicated by the false color, and the phase $\operatorname{arg}\{{D_n}^2\}$ is indicated by the angle of the vector with respect to the real energy axis. Other states with $|{D_n}^2| \leq 0.016$ are represented by smaller black dots. Right panels: Total cross sections and effective cross sections for states in various energy regions in (b) the intrachannel and (d) the full CIS models.}
\end{figure*}

So far, our discussion has been centered on the fundamental $4d^{-1} \epsilon f$ resonance eigenstates of Xe. We now elaborate how the intrinsic electronic structure is mapped onto spectroscopic features in the photoabsorption spectra, starting with the one-photon case relevant to the GDR.

The total one-photon absorption cross section can be constructed from the bottom up, making use of the eigenstates of the complex-scaled $N$-body Hamiltonian \cite{c.buth07b, t.rescigno75a}:
\begin{equation}
\sigma_{\text{tot}} (\omega) = - 4 \pi \alpha \omega \operatorname{Im} \sum_{n} \frac{{D_n}^2}{\omega - E_{n}} \eqqcolon \sum_n \sigma_n (\omega), \label{Eq:Cx}
\end{equation}
where $\alpha$ is the fine structure constant, $D_{n} \coloneqq (\Phi_{n} | \hat{D}_z | \Phi^{\text{HF}}_0)$ is the dipole transition matrix element from the initial state $|\Phi^{\text{HF}}_0)$ to an excited state $|\Phi_{n})$ along the polarization axis $z$, and $\sigma_n (\omega)$ defines the corresponding individual cross section. Eq.~(\ref{Eq:Cx}) enables one to dissect the total cross section into the contributions of various final-state components. More importantly, thanks to complex scaling, each resonance state has a clear-cut contribution, since it is associated with one eigenfunction of the scaled Hamiltonian, rather than with a collection of continuum states of the unscaled Hamiltonian.

The total cross sections in the intrachannel and the full CIS models obtained using this time-independent method [see $\sigma_{\text{tot}} (\omega)$ in Figs.~\ref{Fig:OnePhotonCx}(b) and \ref{Fig:OnePhotonCx}(d)] are in excellent quantitative agreement with those evaluated using a time-dependent CIS method without complex scaling \cite{d.krebs14a}. This verifies that SES does not perturb the wave functions in the physical inner region, where photoabsorption and particle-hole interactions happen. Notice that our intrachannel $\sigma_{\text{tot}} (\omega)$ bears resemblance to Cooper's one-particle spectrum \cite{j.cooper64a}. Also, there is reasonable quantitative agreement between the full CIS $\sigma_{\text{tot}} (\omega)$ and the experimental data \cite{d.krebs14a}.

Figure \ref{Fig:OnePhotonCx}(a) displays the distribution of the squared dipole matrix elements ${D_n}^2$ in polar form for the intrachannel eigenstates. For clarity, only those with $|{D_n}^2| > 0.016$ are shown. The dipole strengths cluster around two separate regions $G_{1}$ and $G_{2}$ in the complex energy plane. We thereupon analyze the effective cross section for all the states in each region, where their individual $\sigma_n(\omega)$ overlap notably and undergo strong constructive or destructive interferences depending on the relative dipole phases. In the first region $G_{1}$ lie only the exposed intrachannel resonances. As their ${D_n}^2$ are large in amplitude and nearly identical in phase, their $\sigma_n$ interfere constructively and bring a net contribution $\sigma_{G_{1}}(\omega)$ that governs the narrow peak at 80 eV in $\sigma_{\text{tot}}(\omega)$ [Fig.~\ref{Fig:OnePhotonCx}(b)]. In the second region $G_{2}$ are the continuum states with $\operatorname{Im} \{ E \} \approx -30 \ \text{eV}$. Owing to the rapid phase variation in ${D_n}^2$, the $\sigma_n$ of those states interfere destructively. Accordingly, the effective $\sigma_{G_{2}}(\omega)$, which implicitly contains the contributions from the unexposed intrachannel resonances, practically plays no part in $\sigma_{\text{tot}}$. All the other states with $|{D_n}^2| \leq 0.016$ cause the minute difference between $\sigma_{G_{1}}+\sigma_{G_{2}}$ and $\sigma_{\text{tot}}$.

Figure \ref{Fig:OnePhotonCx}(c) is the distribution of the ${D_n}^2$ for the eigenstates in the full CIS model. Similar to the intrachannel case, the dipole strengths here are also concentrated in two regions. The first region $G_{1}$ encompasses $R_{1-1}$ and the neighboring continuum states. If one assigns a net transition dipole to the continuum states, it roughly has half of the amplitude and points in the opposite direction in comparison to that of $R_{1-1}$. Due to this destructive interference, all the states in $G_{1}$ jointly produce a weak asymmetric background $\sigma_{G_{1}}(\omega)$ in $\sigma_{\text{tot}}(\omega)$ [Fig.~\ref{Fig:OnePhotonCx}(d)]. In the second region $G_{2}$ resides merely one eigenstate, $R_{1-2}$. For $R_{1-2}$ has a large transition dipole and is very isolated, its own characteristic feature $\sigma_{G_{2}}(\omega)$ stands out without much interference and provides the major contribution to the broad hump around 100 eV in $\sigma_{\text{tot}}$. The discrepancy between $\sigma_{G_{1}}+\sigma_{G_{2}}$ and $\sigma_{\text{tot}}$ mostly stems from the continuum states with $\operatorname{Im} \{ E \} \lesssim -30 \ \text{eV}$.

At this stage, it becomes evident that the linear response of Xe does not reveal the full landscape of the dipole-accessible electronic structure in the $4d$ continuum. In experimental \cite{d.ederer64a, a.lukirskii64a} and conventional theoretical \cite{m.amusia90b, m.amusia00a, c.brechignac94a, j.connerade87a, a.starace82a, j.cooper64a, m.amusia67a, w.brandt67a, a.starace70a, g.wendin73c, a.zangwill80a, z.altun88a} studies, resonance states are known in an implicit, top-down manner---they are inferred from features in the photoabsorption spectrum. Such indirectness explains why the dipole-allowed resonance substructure is unbeknownst to linear spectroscopy. Because of the interferences among overlapping eigenstates, only the exposed resonance group is visible in the intrachannel total one-photon cross section, and only the broader dipole sub-resonance $R_{1-2}$ shows up as the GDR in the full CIS cross section.

In both the one-particle and the many-body one-photon absorption spectra, there is only one smooth, structureless resonance-like feature. As such, the energy up-shift and the broadening of the maximum in $\sigma_{\text{tot}}(\omega)$ upon the inclusion of multielectron interactions have long been interpreted, since the seminal work of Cooper in 1964 \cite{j.cooper64a}, as a correlation-induced modification of the Siegert energy of one single resonance state \cite{j.cooper64a, m.amusia67a, w.brandt67a, a.starace70a, g.wendin73c, a.zangwill80a}. Per contra, we find that these spectroscopic changes actually arise from switching the visibilities of two distinct resonance groups without any adiabatic connection. Whereas interchannel couplings suppress the fingerprint of the exposed intrachannel resonances by moving $R_{1-1}$ downward in the energy plane and introducing destructive interferences with nearby continuum states, they enhance the feature of the unexposed ones by moving $R_{1-2}$ horizontally away from the continuum and eliminating destructive interferences.

\subsection{\label{Subsec:ThreePhotonCx}Three-photon--two-color proposal for probing the collective octupole resonance}
\begin{figure}
\includegraphics[angle=270]{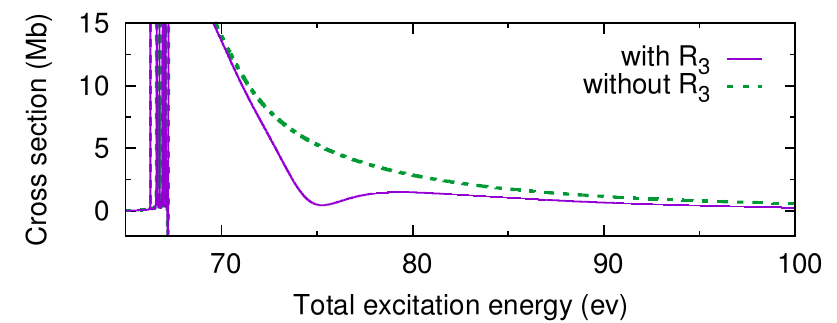}
\caption{\label{Fig:ThreePhotonCx}One-photon cross sections with and without the contribution of $R_3$ as a function of the total three-photon excitation energy relative to the Xe ground state. The initial state is presumed the lowest $4d^{-1} n d (^1D)$ Rydberg state.}
\end{figure}

We have analyzed the optical response attributable to the $4d$ dipole resonances of Xe. We now present an effective three-photon--two-color scheme that aims to unveil the response of the collective octupole resonance $R_3$. Our proposal may be realized with a combination of XUV-FELs \cite{j.feldhaus10a, e.allaria10a} and lower-order HHG sources \cite{t.sekikawa04a}. This two-color scheme has the advantage of its target selectivity: it has been shown that, on a mean-field level, a simpler three-photon--one-color scheme excites the Xe ground state predominantly to the final states with $4d^{-1} \epsilon f (^1P)$, but not to those with $4d^{-1} \epsilon f (^1F)$ \cite{l.pi14a}.

In the first step of the proposed scheme, the ground state is promoted to the lowest $4d^{-1} n d (^1D)$ Rydberg state at 64.8 eV through two-photon resonant excitation. This requires an intense XUV source with a photon energy of 32.4 eV and a sub-eV bandwidth in order to avoid the excitation of other bound states with even parity. In the second step, the $4d^{-1} n d (^1D)$ state absorbs one photon with a different color and goes to $R_3$ [$4d^{-1}\epsilon f (^1F)$]. This necessitates a weak source at a photon energy of 9.5 eV in the VUV. The target specificity of our scheme is rationalized by the fact that the lowest $4d^{-1} n d (^1D)$ state has the strongest dipole transition matrix element to $R_3$ as compared to all the other states lying above the $4d$ ionization threshold.

As a proof of concept, we assume that the first excitation step generates a system in a pure state of the lowest $4d^{-1} n d (^1D)$ state. The total one-photon absorption cross section corresponding to the second step can then be calculated by substituting the $4d^{-1} n d (^1D)$ wave function for $| \Phi^{\text{HF}}_0)$ in Eq.~(\ref{Eq:Cx}). For the results shown below, we utilize another set of SES parameters. The reason and the numerical values are provided in Appendix \ref{Sec:NumDtl}.

Figure \ref{Fig:ThreePhotonCx} plots the one-photon cross sections with and without the contribution of $R_3$ as a function of the total three-photon excitation energy relative to the Xe ground state. Comparison between the two spectra verifies that the window resonance close to 75 eV originates precisely from the target $R_3$.

Note that as the current XUV-FELs \cite{j.feldhaus10a, e.allaria10a} have a typical pulse duration longer than the Auger decay time of $4d^{-1}$ \cite{m.jurvansuu01a}, the XUV and VUV pulses have to overlap in time. Experimentally, this requires the two pulses to be synchronized with femtosecond accuracy. Photoelectron spectroscopy can help to disentangle the signal of $R_3$ from the background due to XUV or VUV absorption of the valence electrons. The kinetic energy of the photoelectron associated with the final state $R_3$ is 6.8 eV. Close to this energy, XUV one-photon ionization of the $5s$ subshell can yield a photoelectron carrying a kinetic energy of 9.1 eV. By further measuring the angular-resolved photoelectron distribution, it is feasible to separate the contribution of the target (with an $f$-wave character) from that of the background (with a $p$-wave character). On the theory side, the temporal overlap of the two pulses means that further studies on the cross section that take into account the effects of virtual excitations are expected to improve the quantitative details presented here.

\section{\label{Sec:Conclusion}Conclusion}
Summarizing, this paper tackles the collective resonances in the $4d$ subshell of Xe at the most fundamental level of the correlated electronic structure. Direct knowledge of the resonance eigenstates from first principles is made possible by employing the SES procedure within the wave-function--based many-body CIS theory. By explicitly tracking the adiabatic evolution of the resonance states, we demonstrate the diverse role of electronic correlations in the formation of various types of $4d^{-1} \epsilon f$ collective resonances. By examining the one-photon cross section in a bottom-up, state-resolved fashion, we show that linear spectroscopy reflects only partial information of the dipole-allowed resonance-state substructure as a result of interference effects. Combining the results of the adiabatic trajectories and the cross section analysis, we find that, when going from Cooper's one-particle spectrum to the experimental data, the spectroscopic changes cannot be obtained by adding correlations to one resonance eigenstate; they are signatures of swapping the visibilities of two individual resonances without any adiabatic connection. Moreover, we predict the existence of atomic multipole plasmons, which holds promise for accessing the dynamics of Xe under the intertwined effects of collectivity and nonlinearity, a barely explored territory \cite{m.amusia00a}.

Finally, we point out that a recent experiment has applied attosecond pump-probe techniques to detect shape resonances in molecular valence orbitals \cite{m.huppert16a}. With a newly achieved experimental timing capability down to a few attoseconds \cite{m.ossiander17a}, it may be well expected that experiments of this kind will offer a test ground for our results in real time. Beginning with the prototypical collective excitations in atomic systems, the insights and methodology in this work pave the way towards a deeper understanding of the collective response of matter to light---from the linear to the nonlinear regime.

\begin{acknowledgments}
We thank G.~Wendin, O.~Vendrell, M.~Ossiander, and M.~Schultze for helpful discussions. S.P.~is funded by the Alexander von Humboldt Foundation and by the NSF through a grant to ITAMP.
\end{acknowledgments}

\appendix
\section{\label{Sec:NumDtl}Computational details}
In this appendix, we present the numerical details of our calculations.

\begin{itemize}
\item Active ionization channels $\mathbf{i}$ \cite{s.pabst12b} include all the orbitals in the $4d$, $5s$, and $5p$ subshells. The orbital energies of $4d$ \cite{a.thompson09a, m.jurvansuu01a}, $5s$ \cite{a.kramida14a}, and $5p$ \cite{a.kramida14a} are slightly adjusted by hand to match the experimental values.

\item We follow the SES path in Refs.~\cite{y.chen15a, s.pabst16b} for the analytic continuation of the electron radial coordinate into the complex plane. For Figs.~\ref{Fig:Ses}, \ref{Fig:SesEvo}, and \ref{Fig:OnePhotonCx}, the scaling starts at $r_0 = 8 \ \text{a.u.}$, the rotation angle $\theta$ is $36^\circ$, and the smoothing parameter $\lambda$ is 1 a.u.. These parameters are required to expose all the physically relevant resonance poles and to prevent the continuum branching artifact \cite{u.riss93a}. In order to ensure the Xe ground state is unperturbed by the scaling, we examine the results by varying the SES parameters over a sensible range, e.g., $r_0$ between 8 and 20 a.u.~and $\theta$ between 35 and $44.5^\circ$. By doing so, we estimate an uncertainty of 0.5 eV for the Siegert energies shown in Table \ref{Tab:SesEngy}.

\item The second group of intrachannel resonances (with $\operatorname{Re}\{E\} \approx 70 \ \text{eV}, \ \operatorname{Im}\{E\} \approx -30 \ \text{eV}$ as indicated by the black crosses in Fig.~\ref{Fig:SesEvo}) cannot be exposed using the above $\theta = 36^\circ$. Their Siegert energies are calculated with an even larger scaling angle $\theta = 44^\circ$. The reason why we keep them unexposed is because such a large scaling angle, although it has little influence on the Siegert energies, does impair the quality of the resonance and continuum wave functions and thus the quality of the photoabsorption cross sections.

\item The SES parameter $r_0 = 8 \ \text{a.u.}$ used for Figs.~\ref{Fig:Ses}, \ref{Fig:SesEvo}, and \ref{Fig:OnePhotonCx} perturbs the lowest $4d^{-1} n d (^1D)$ Rydberg state in the three-photon--two-color scheme of Sec.~\ref{Subsec:ThreePhotonCx}. For Fig.~\ref{Fig:ThreePhotonCx}, we therefore use a more moderate $r_0 = 20 \ \text{a.u.}$. This also makes sure that the atomic multipole plasmons predicted in this work are independent of the numerical parameters and are thus not numerical artifacts.

\item We adopt the nonuniform radial grid in Ref.~\cite{l.greenman10a}: the radial grid size $r_{\text{max}}$ is 250 a.u.; the mapping parameter $\zeta$ is 1 a.u.; the number of grid points $N$ is 1800. The radial basis functions are constructed using the finite-element discrete-variable representation described in Ref.~\cite{s.pabst16b}.

\item The maximum orbital angular momentum quantum number $l_{\text{max}}$ \cite{s.pabst12b} is 4.

\item Numerical diagonalization of the $N$-electron Hamiltonian is done by the Arnoldi iteration using the \textsc{ARPACK} library \cite{d.sorensen96a}. An initial random vector is used to launch the iteration.

\item To construct the total one-photon absorption cross sections $\sigma_{\text{tot}} (\omega)$ in Figs.~\ref{Fig:OnePhotonCx}(b) and \ref{Fig:OnePhotonCx}(d), we include the contributions of all the eigenstates with an excitation energy between 60 and 200 eV. We evaluate the dipole transition matrix elements $D_{n}$ using the dipole operator in the velocity form \cite{n.rohringer06a}. As shown by a CIS wave-packet calculation without complex scaling, the velocity form predicts a one-photon absorption spectrum in closer agreement with the experimental data \cite{d.krebs13a, d.krebs14a}.
\end{itemize}

\section{\label{Sec:OneBodyRes}Properties of the double-pole structure in the one-particle limit}
The double-pole structure in our noninteracting model, i.e., the exposed and the unexposed intrachannel resonance groups in Fig.~\ref{Fig:SesEvo}, comes as a surprising result. Since destructive interferences forbid the unexposed resonance group to be seen in Cooper's one-particle spectrum \cite{j.cooper64a}, its presence was previously unknown. In addition, it is rather unusual that a potential can support various resonance groups with distinct decay widths yet similar excitation energies \cite{n.moiseyev98b, r.santra02a}. Following this observation, we hence investigate the properties of the one-particle potential necessary for the existence of such a double-pole structure. This is also a first step towards understanding the different roles collectivity plays in forming $R_{1-1}$ and $R_{1-2}$, respectively.

In the main text, we ascribe the appearance of the double-pole structure to the nonlocal nature of the ionic potential. In this appendix, we provide numerical evidence to support this statement.

\subsection{\label{Subsec:Method}Methodology}
We do not explicitly construct the effective one-particle potential here. Nevertheless, the properties of the potential can be tuned implicitly by manipulating the intrachannel-coupling Coulomb matrix elements \cite{s.pabst12b, s.pabst13a}. Within CIS, the part of the $N$-body Hamiltonian that contains the residual electron-ion interactions beyond the description of the mean-field HF potential is \cite{s.pabst12b, s.pabst13a}:
\begin{align} 
\hat{H}_{\text{e-e}} &= \frac{1}{2} \sum_{\substack{n, n'=1 \\ n \neq n' }}^{N} \frac{1}{\left| \hat{\mathbf{r}}_n - \hat{\mathbf{r}}_{n'} \right|}  - \sum_{n=1}^N \hat{V}^{\text{HF}}(\hat{\mathbf{r}}_n). \label{Eq:ResidualH}
\end{align}
In the intrachannel model, we consider only the matrix elements of the type $( \Phi^\mathbf{b}_\mathbf{i} | \hat{H}_{\text{e-e}} | \Phi^\mathbf{a}_\mathbf{i} )$. This one-body part of $\hat{H}_{\text{e-e}}$ corresponds to the picture of an excited electron moving in the presence of an attractive ionic potential \cite{s.pabst12b, s.pabst13a}.

In the following, we analyze two scenarios that impose further restrictions on the intrachannel-coupling terms. Because the behaviors of different $4d_{\pm m}$ ionization channels are much alike, we only show the results for the $4d_{\pm 1}$ channel. The calculations are done using a scaling angle of $\theta = 44.5^\circ$, which can directly expose both resonance groups in the intrachannel model. This is also the largest possible scaling angle for our SES procedure: any scaling angle bigger than that will lead to noticeable perturbation of the Xe ground state.

\subsection{\label{Subsec:CmprHF}Importance of the ionic potential}
\begin{figure}
\includegraphics[angle=270]{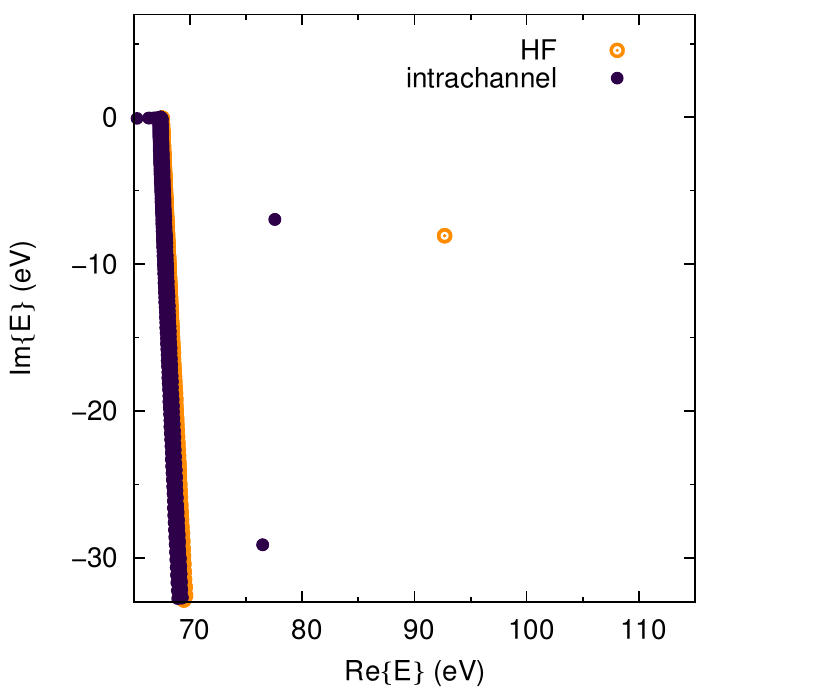}
\caption{\label{Fig:CmprHF}Comparison between the complex energy spectrum in the HF model and that in the intrachannel model.}
\end{figure}

In the first scenario, we simply switch off the residual intrachannel Coulomb interactions, so the one-particle potential reduces to the HF potential of a neutral Xe atom.

Figure \ref{Fig:CmprHF} shows the comparison between the SES energy spectrum in the HF model and that in the intrachannel model. The intrachannel model gives rise to the double-pole structure, while the HF model predicts only one $4d_{\pm 1}^{-1} \epsilon f_{\pm 1}$ resonance. Apparently, the double-pole structure, or more specifically the shorter-lived resonance, cannot be captured by the short-range potential of neutral $\text{Xe}$ and requires the long-range potential of $\text{Xe}^{+}$. In the next step, we divide the ionic potential into various components and then discuss their individual impacts.

\subsection{\label{Subsec:MtplSwitch}Importance of the exchange ionic potential}
To tackle the Coulomb integrals in Eq.~(\ref{Eq:ResidualH}), we perform the multipole expansion \cite{l.greenman10a}:
\begin{widetext}
\begin{align}
\hat{r}_{1,2}^{-1} \eqqcolon \frac{1}{\left| \hat{\mathbf{r}}_1 (r_1, \theta_1, \phi_1) - \hat{\mathbf{r}}_2 (r_2, \theta_2, \phi_2) \right|} = \sum_{L=0}^{\infty} \frac{4\pi}{2L+1} \frac{r^L_<}{r^{L+1}_>} \sum_{M = -L}^{L} Y_{L,M}^*(\theta_1, \phi_1) Y_{L,M}(\theta_2, \phi_2). \label{Eq:MtplExp}
\end{align}
\end{widetext}
For explanations of the notations, please see Ref.~\cite{l.greenman10a}.

In the intrachannel model, the direct Coulomb matrix elements read \cite{n.rohringer06a, s.pabst12b}:
\begin{align}
v_{(\mathbf{b}, \mathbf{i}, \mathbf{a}, \mathbf{i})} &= \Big( \phi_{\mathbf{b}}(\hat{\mathbf{r}}_1) \phi_{\mathbf{i}} (\hat{\mathbf{r}}_2) \Big| \hat{r}_{1,2}^{-1} \Big| \phi_{\mathbf{a}}(\hat{\mathbf{r}}_1) \phi_{\mathbf {i}} (\hat{\mathbf{r}}_2) \Big), \label{Eq:CoulombDic}
\end{align}
and the exchange terms read \cite{n.rohringer06a, s.pabst12b}:
\begin{align}
v_{(\mathbf{b}, \mathbf{i}, \mathbf{i}, \mathbf{a})} &= \Big( \phi_{\mathbf{b}}(\hat{\mathbf{r}}_1) \phi_{\mathbf{i}} (\hat{\mathbf{r}}_2) \Big| \hat{r}_{1,2}^{-1} \Big| \phi_{\mathbf{i}}(\hat{\mathbf{r}}_1) \phi_{\mathbf{a}}(\hat{\mathbf{r}}_2) \Big). \label{Eq:CoulombExc}
\end{align}
In these expressions, $\phi_{\mathbf{i}}$ denotes a hole orbital occupied in the HF ground state, and $\phi_{\mathbf{a}}$ and $\phi_{\mathbf{b}}$ symbolize virtual orbitals of the excited electron \cite{n.rohringer06a}. The angular parts of the orbitals are expressed in terms of spherical harmonics \cite{l.greenman10a}.

Let us simply analyze the angular parts of the matrix elements. Inserting Eq.~(\ref{Eq:MtplExp}) into Eqs.~(\ref{Eq:CoulombDic}) and (\ref{Eq:CoulombExc}) respectively, we obtain the following relationships:

\begin{align}
&\text{Angular part of the $L$-th order term of} \ v_{(\mathbf{b}, \mathbf{i}, \mathbf{a}, \mathbf{i})} \nonumber \\
&\propto \Bra{Y_{l_b, m_b}} Y_{L,M}^* \Ket{Y_{l_a, m_a}} \Bra{Y_{l_i, m_i}} Y_{L,M} \Ket{Y_{l_i, m_i}}, \label{Eq:CoulombDicAng}
\end{align}
and
\begin{align}
&\text{Angular part of the $L$-th order term of} \ v_{(\mathbf{b}, \mathbf{i}, \mathbf{i}, \mathbf{a})} \nonumber \\
&\propto \Bra{Y_{l_b, m_b}} Y_{L,M}^* \Ket{Y_{l_i, m_i}} \Bra{Y_{l_i, m_i}} Y_{L,M} \Ket{Y_{l_a, m_a}}. \label{Eq:CoulombExcAng}
\end{align}
For an $f$-wave intrachannel resonance of a specific $4d_{m_i}$ channel, $l_i = 2$, $l_a = l_b = 3$, and $m_i = m_a = m_b$. According to the addition of angular momenta, we only have to consider the cases with $M = 0$. More importantly, due to the parity of spherical harmonics, the direct terms are nonzero only for an even $L$, and the exchange terms are nonzero only for an odd $L$. Thus, as one carries out a multipole expansion of the intrachannel Coulomb matrix elements, not only does one divide the photoelectron-ion interactions based on the angular features, but one also separates the local direct part of the ionic potential from the nonlocal exchange part.

\begin{figure}
\includegraphics[angle=270]{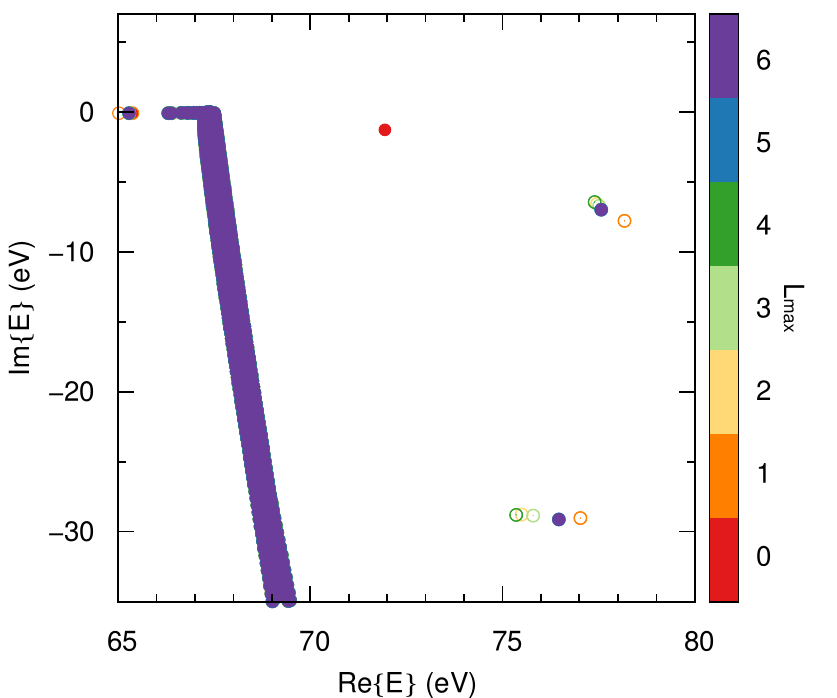}
\caption{\label{Fig:MtplSwitch}Evolution of the complex energy spectrum when including up to the $L_{\text{max}}$-th order term in the multipole expansion of the intrachannel-coupling Coulomb matrix elements. The parameter $L_{\text{max}}$ is indicated by the false color. The double-pole structure emerges upon $L_{\text{max}} = 1$.}
\end{figure}

Figure \ref{Fig:MtplSwitch} illustrates the emergence of the double-pole structure in the intrachannel model when including up to the $L_{\text{max}}$-th order term in the multipole expansion of the intrachannel Coulomb matrix elements. With exclusively the monopolar $L=0$ term, the leading order contribution of the direct interactions, there is only the longer-lived uncorrelated resonance but not the shorter-lived one. However, as soon as one adds in the dipolar $L = 1$ term, the leading order contribution of the exchange interactions, the double poles with almost the same $\operatorname{Re}\{E\}$ but very different $\operatorname{Im}\{E\}$ immediately spring up. Putting in even higher order terms only causes small perturbations of the Siegert energies of the two resonances, which are converged for $L \geq  5$.

As substantiated by Fig.~\ref{Fig:MtplSwitch}, the appearance of the double-pole structure in the one-particle limit is an immediate consequence of the exchange ionic potential sensed by the photoelectron, which enters the intrachannel Coulomb matrix elements to leading order through a dipolar term. Due to the sensitivity to such nonlocality, the shorter-lived resonance cannot be trivially explained by the notion of shape resonances, which portrays an outgoing electron as being trapped by a local angular momentum barrier \cite{a.starace82a}. As a final remark, we point out that the monopolar component of the ionic potential, which represents a simple central field $V(r)$, is insufficient for the description of the double-pole structure.

\bibliography{Chen_JPB}

\begin{thebibliography}{60}%
\makeatletter
\providecommand \@ifxundefined [1]{%
 \@ifx{#1\undefined}
}%
\providecommand \@ifnum [1]{%
 \ifnum #1\expandafter \@firstoftwo
 \else \expandafter \@secondoftwo
 \fi
}%
\providecommand \@ifx [1]{%
 \ifx #1\expandafter \@firstoftwo
 \else \expandafter \@secondoftwo
 \fi
}%
\providecommand \natexlab [1]{#1}%
\providecommand \enquote  [1]{``#1''}%
\providecommand \bibnamefont  [1]{#1}%
\providecommand \bibfnamefont [1]{#1}%
\providecommand \citenamefont [1]{#1}%
\providecommand \href@noop [0]{\@secondoftwo}%
\providecommand \href [0]{\begingroup \@sanitize@url \@href}%
\providecommand \@href[1]{\@@startlink{#1}\@@href}%
\providecommand \@@href[1]{\endgroup#1\@@endlink}%
\providecommand \@sanitize@url [0]{\catcode `\\12\catcode `\$12\catcode
  `\&12\catcode `\#12\catcode `\^12\catcode `\_12\catcode `\%12\relax}%
\providecommand \@@startlink[1]{}%
\providecommand \@@endlink[0]{}%
\providecommand \url  [0]{\begingroup\@sanitize@url \@url }%
\providecommand \@url [1]{\endgroup\@href {#1}{\urlprefix }}%
\providecommand \urlprefix  [0]{URL }%
\providecommand \Eprint [0]{\href }%
\providecommand \doibase [0]{http://dx.doi.org/}%
\providecommand \selectlanguage [0]{\@gobble}%
\providecommand \bibinfo  [0]{\@secondoftwo}%
\providecommand \bibfield  [0]{\@secondoftwo}%
\providecommand \translation [1]{[#1]}%
\providecommand \BibitemOpen [0]{}%
\providecommand \bibitemStop [0]{}%
\providecommand \bibitemNoStop [0]{.\EOS\space}%
\providecommand \EOS [0]{\spacefactor3000\relax}%
\providecommand \BibitemShut  [1]{\csname bibitem#1\endcsname}%
\let\auto@bib@innerbib\@empty
\bibitem [{\citenamefont {Bloch}(1933)}]{f.bloch33a}%
  \BibitemOpen
  \bibfield  {author} {\bibinfo {author} {\bibfnamefont {F.}~\bibnamefont
  {Bloch}},\ }\href@noop {} {\bibfield  {journal} {\bibinfo  {journal} {Z.
  Phys.}\ }\textbf {\bibinfo {volume} {81}},\ \bibinfo {pages} {363} (\bibinfo
  {year} {1933})}\BibitemShut {NoStop}%
\bibitem [{\citenamefont {Amusia}(1990)}]{m.amusia90b}%
  \BibitemOpen
  \bibfield  {author} {\bibinfo {author} {\bibfnamefont {M.~Ya.}\ \bibnamefont
  {Amusia}},\ }\href@noop {} {\emph {\bibinfo {title} {{Atomic Photoeffect}}}}\
  (\bibinfo  {publisher} {Plenum},\ \bibinfo {address} {New York},\ \bibinfo
  {year} {1990})\BibitemShut {NoStop}%
\bibitem [{\citenamefont {Amusia}\ and\ \citenamefont
  {Connerade}(2000)}]{m.amusia00a}%
  \BibitemOpen
  \bibfield  {author} {\bibinfo {author} {\bibfnamefont {M.~Ya.}\ \bibnamefont
  {Amusia}}\ and\ \bibinfo {author} {\bibfnamefont {J.~P.}\ \bibnamefont
  {Connerade}},\ }\href@noop {} {\bibfield  {journal} {\bibinfo  {journal}
  {Rep. Prog. Phys.}\ }\textbf {\bibinfo {volume} {63}},\ \bibinfo {pages} {41}
  (\bibinfo {year} {2000})}\BibitemShut {NoStop}%
\bibitem [{\citenamefont {Br{\'{e}}chignac}\ and\ \citenamefont
  {Connerade}(1994)}]{c.brechignac94a}%
  \BibitemOpen
  \bibfield  {author} {\bibinfo {author} {\bibfnamefont {C.}~\bibnamefont
  {Br{\'{e}}chignac}}\ and\ \bibinfo {author} {\bibfnamefont {J.~P.}\
  \bibnamefont {Connerade}},\ }\href@noop {} {\bibfield  {journal} {\bibinfo
  {journal} {J. Phys. B At. Mol. Opt. Phys.}\ }\textbf {\bibinfo {volume}
  {27}},\ \bibinfo {pages} {3795} (\bibinfo {year} {1994})}\BibitemShut
  {NoStop}%
\bibitem [{\citenamefont {Connerade}\ \emph {et~al.}(1987)\citenamefont
  {Connerade}, \citenamefont {Esteva},\ and\ \citenamefont
  {Karnatak}}]{j.connerade87a}%
  \BibitemOpen
  \bibinfo {editor} {\bibfnamefont {J.~P.}\ \bibnamefont {Connerade}}, \bibinfo
  {editor} {\bibfnamefont {J.~M.}\ \bibnamefont {Esteva}}, \ and\ \bibinfo
  {editor} {\bibfnamefont {R.~C.}\ \bibnamefont {Karnatak}},\ eds.,\ \href@noop
  {} {\emph {\bibinfo {title} {{Giant Resonances in Atoms, Molecules, and
  Solids}}}}\ (\bibinfo  {publisher} {Plenum},\ \bibinfo {address} {New York},\
  \bibinfo {year} {1987})\BibitemShut {NoStop}%
\bibitem [{\citenamefont {Starace}(1982)}]{a.starace82a}%
  \BibitemOpen
  \bibfield  {author} {\bibinfo {author} {\bibfnamefont {A.~F.}\ \bibnamefont
  {Starace}},\ }in\ \href@noop {} {\emph {\bibinfo {booktitle} {Encyclopedia of Physics: Corpuscles and
Radiation in Matter I}}},\ Vol.~\bibinfo {volume} {31},\ \bibinfo {editor} {edited by\
  \bibinfo {editor} {\bibfnamefont {W.}~\bibnamefont {Mehlhorn}}}\ (\bibinfo
  {publisher} {Springer-Verlag},\ \bibinfo {address} {Berlin},\ \bibinfo {year}
  {1982})\ p.~\bibinfo {pages} {1}\BibitemShut {NoStop}%
\bibitem [{\citenamefont {Cooper}(1964)}]{j.cooper64a}%
  \BibitemOpen
  \bibfield  {author} {\bibinfo {author} {\bibfnamefont {J.~W.}\ \bibnamefont
  {Cooper}},\ }\href@noop {} {\bibfield  {journal} {\bibinfo  {journal} {Phys.
  Rev. Lett.}\ }\textbf {\bibinfo {volume} {13}},\ \bibinfo {pages} {762}
  (\bibinfo {year} {1964})}\BibitemShut {NoStop}%
\bibitem [{\citenamefont {Amusia}\ \emph {et~al.}(1967)\citenamefont {Amusia},
  \citenamefont {Cherepkov},\ and\ \citenamefont {Sheftel}}]{m.amusia67a}%
  \BibitemOpen
  \bibfield  {author} {\bibinfo {author} {\bibfnamefont {M.~Ya.}\ \bibnamefont
  {Amusia}}, \bibinfo {author} {\bibfnamefont {N.~A.}\ \bibnamefont
  {Cherepkov}}, \ and\ \bibinfo {author} {\bibfnamefont {S.~I.}\ \bibnamefont
  {Sheftel}},\ }\href@noop {} {\bibfield  {journal} {\bibinfo  {journal} {Phys.
  Lett. A}\ }\textbf {\bibinfo {volume} {24}},\ \bibinfo {pages} {541}
  (\bibinfo {year} {1967})}\BibitemShut {NoStop}%
\bibitem [{\citenamefont {Brandt}\ \emph {et~al.}(1967)\citenamefont {Brandt},
  \citenamefont {Eder},\ and\ \citenamefont {Lundqvist}}]{w.brandt67a}%
  \BibitemOpen
  \bibfield  {author} {\bibinfo {author} {\bibfnamefont {W.}~\bibnamefont
  {Brandt}}, \bibinfo {author} {\bibfnamefont {L.}~\bibnamefont {Eder}}, \ and\
  \bibinfo {author} {\bibfnamefont {S.}~\bibnamefont {Lundqvist}},\ }\href@noop
  {} {\bibfield  {journal} {\bibinfo  {journal} {J. Quant. Spectrosc. Radiat.
  Transf.}\ }\textbf {\bibinfo {volume} {7}},\ \bibinfo {pages} {185} (\bibinfo
  {year} {1967})}\BibitemShut {NoStop}%
\bibitem [{\citenamefont {Starace}(1970)}]{a.starace70a}%
  \BibitemOpen
  \bibfield  {author} {\bibinfo {author} {\bibfnamefont {A.~F.}\ \bibnamefont
  {Starace}},\ }\href@noop {} {\bibfield  {journal} {\bibinfo  {journal} {Phys.
  Rev. A}\ }\textbf {\bibinfo {volume} {2}},\ \bibinfo {pages} {118} (\bibinfo
  {year} {1970})}\BibitemShut {NoStop}%
\bibitem [{\citenamefont {Wendin}(1973)}]{g.wendin73c}%
  \BibitemOpen
  \bibfield  {author} {\bibinfo {author} {\bibfnamefont {G.}~\bibnamefont
  {Wendin}},\ }\href@noop {} {\bibfield  {journal} {\bibinfo  {journal} {J.
  Phys. B At. Mol. Opt. Phys.}\ }\textbf {\bibinfo {volume} {6}},\ \bibinfo
  {pages} {42} (\bibinfo {year} {1973})}\BibitemShut {NoStop}%
\bibitem [{\citenamefont {Zangwill}\ and\ \citenamefont
  {Soven}(1980)}]{a.zangwill80a}%
  \BibitemOpen
  \bibfield  {author} {\bibinfo {author} {\bibfnamefont {A.}~\bibnamefont
  {Zangwill}}\ and\ \bibinfo {author} {\bibfnamefont {P.}~\bibnamefont
  {Soven}},\ }\href@noop {} {\bibfield  {journal} {\bibinfo  {journal} {Phys.
  Rev. A}\ }\textbf {\bibinfo {volume} {21}},\ \bibinfo {pages} {1561}
  (\bibinfo {year} {1980})}\BibitemShut {NoStop}%
\bibitem [{\citenamefont {Ederer}(1964)}]{d.ederer64a}%
  \BibitemOpen
  \bibfield  {author} {\bibinfo {author} {\bibfnamefont {D.~L.}\ \bibnamefont
  {Ederer}},\ }\href@noop {} {\bibfield  {journal} {\bibinfo  {journal} {Phys.
  Rev. Lett.}\ }\textbf {\bibinfo {volume} {13}},\ \bibinfo {pages} {760}
  (\bibinfo {year} {1964})}\BibitemShut {NoStop}%
\bibitem [{\citenamefont {Lukirskii}\ \emph {et~al.}(1964)\citenamefont
  {Lukirskii}, \citenamefont {Brytov},\ and\ \citenamefont
  {Zimkina}}]{a.lukirskii64a}%
  \BibitemOpen
  \bibfield  {author} {\bibinfo {author} {\bibfnamefont {A.~P.}\ \bibnamefont
  {Lukirskii}}, \bibinfo {author} {\bibfnamefont {I.~A.}\ \bibnamefont
  {Brytov}}, \ and\ \bibinfo {author} {\bibfnamefont {T.~M.}\ \bibnamefont
  {Zimkina}},\ }\href@noop {} {\bibfield  {journal} {\bibinfo  {journal} {Opt.
  Spectrosc.}\ }\textbf {\bibinfo {volume} {17}},\ \bibinfo {pages} {234}
  (\bibinfo {year} {1964})}\BibitemShut {NoStop}%
\bibitem [{\citenamefont {Shiner}\ \emph {et~al.}(2011)\citenamefont {Shiner},
  \citenamefont {Schmidt}, \citenamefont {Trallero-Herrero}, \citenamefont
  {W\"{o}rner}, \citenamefont {Patchkovskii}, \citenamefont {Corkum},
  \citenamefont {Kieffer}, \citenamefont {L\'{e}gar\'{e}},\ and\ \citenamefont
  {Villeneuve}}]{a.shiner11a}%
  \BibitemOpen
  \bibfield  {author} {\bibinfo {author} {\bibfnamefont {A.~D.}\ \bibnamefont
  {Shiner}}, \bibinfo {author} {\bibfnamefont {B.~E.}\ \bibnamefont {Schmidt}},
  \bibinfo {author} {\bibfnamefont {C.}~\bibnamefont {Trallero-Herrero}},
  \bibinfo {author} {\bibfnamefont {H.~J.}\ \bibnamefont {W\"{o}rner}},
  \bibinfo {author} {\bibfnamefont {S.}~\bibnamefont {Patchkovskii}}, \bibinfo
  {author} {\bibfnamefont {P.~B.}\ \bibnamefont {Corkum}}, \bibinfo {author}
  {\bibfnamefont {J.-C.}\ \bibnamefont {Kieffer}}, \bibinfo {author}
  {\bibfnamefont {F.}~\bibnamefont {L\'{e}gar\'{e}}}, \ and\ \bibinfo {author}
  {\bibfnamefont {D.~M.}\ \bibnamefont {Villeneuve}},\ }\href {\doibase
  10.1038/nphys1940} {\bibfield  {journal} {\bibinfo  {journal} {Nat. Phys.}\
  }\textbf {\bibinfo {volume} {7}},\ \bibinfo {pages} {464} (\bibinfo {year}
  {2011})}\BibitemShut {NoStop}%
\bibitem [{\citenamefont {Pabst}\ and\ \citenamefont
  {Santra}(2013)}]{s.pabst13b}%
  \BibitemOpen
  \bibfield  {author} {\bibinfo {author} {\bibfnamefont {S.}~\bibnamefont
  {Pabst}}\ and\ \bibinfo {author} {\bibfnamefont {R.}~\bibnamefont {Santra}},\
  }\href {\doibase 10.1103/PhysRevLett.111.233005} {\bibfield  {journal}
  {\bibinfo  {journal} {Phys. Rev. Lett.}\ }\textbf {\bibinfo {volume} {111}},\
  \bibinfo {pages} {233005} (\bibinfo {year} {2013})}\BibitemShut {NoStop}%
\bibitem [{\citenamefont {Richter}\ \emph {et~al.}(2009)\citenamefont
  {Richter}, \citenamefont {Amusia}, \citenamefont {Bobashev}, \citenamefont
  {Feigl}, \citenamefont {Jurani{\'{c}}}, \citenamefont {Martins},
  \citenamefont {Sorokin},\ and\ \citenamefont {Tiedtke}}]{m.richter09a}%
  \BibitemOpen
  \bibfield  {author} {\bibinfo {author} {\bibfnamefont {M.}~\bibnamefont
  {Richter}}, \bibinfo {author} {\bibfnamefont {M.~Ya.}\ \bibnamefont {Amusia}},
  \bibinfo {author} {\bibfnamefont {S.~V.}\ \bibnamefont {Bobashev}}, \bibinfo
  {author} {\bibfnamefont {T.}~\bibnamefont {Feigl}}, \bibinfo {author}
  {\bibfnamefont {P.~N.}\ \bibnamefont {Jurani{\'{c}}}}, \bibinfo {author}
  {\bibfnamefont {M.}~\bibnamefont {Martins}}, \bibinfo {author} {\bibfnamefont
  {A.~A.}\ \bibnamefont {Sorokin}}, \ and\ \bibinfo {author} {\bibfnamefont
  {K.}~\bibnamefont {Tiedtke}},\ }\href {\doibase
  10.1103/PhysRevLett.102.163002} {\bibfield  {journal} {\bibinfo  {journal}
  {Phys. Rev. Lett.}\ }\textbf {\bibinfo {volume} {102}},\ \bibinfo {pages}
  {163002} (\bibinfo {year} {2009})}\BibitemShut {NoStop}%
\bibitem [{\citenamefont {Gerken}\ \emph {et~al.}(2014)\citenamefont {Gerken},
  \citenamefont {Klumpp}, \citenamefont {Sorokin}, \citenamefont {Tiedtke},
  \citenamefont {Richter}, \citenamefont {B{\"{u}}rk}, \citenamefont {Mertens},
  \citenamefont {Jurani{\'{c}}},\ and\ \citenamefont {Martins}}]{n.gerken14a}%
  \BibitemOpen
  \bibfield  {author} {\bibinfo {author} {\bibfnamefont {N.}~\bibnamefont
  {Gerken}}, \bibinfo {author} {\bibfnamefont {S.}~\bibnamefont {Klumpp}},
  \bibinfo {author} {\bibfnamefont {A.~A.}\ \bibnamefont {Sorokin}}, \bibinfo
  {author} {\bibfnamefont {K.}~\bibnamefont {Tiedtke}}, \bibinfo {author}
  {\bibfnamefont {M.}~\bibnamefont {Richter}}, \bibinfo {author} {\bibfnamefont
  {V.}~\bibnamefont {B{\"{u}}rk}}, \bibinfo {author} {\bibfnamefont
  {K.}~\bibnamefont {Mertens}}, \bibinfo {author} {\bibfnamefont
  {P.}~\bibnamefont {Jurani{\'{c}}}}, \ and\ \bibinfo {author} {\bibfnamefont
  {M.}~\bibnamefont {Martins}},\ }\href {\doibase
  10.1103/PhysRevLett.112.213002} {\bibfield  {journal} {\bibinfo  {journal}
  {Phys. Rev. Lett.}\ }\textbf {\bibinfo {volume} {112}},\ \bibinfo {pages}
  {213002} (\bibinfo {year} {2014})}\BibitemShut {NoStop}%
\bibitem [{\citenamefont {Mazza}\ \emph {et~al.}(2015)\citenamefont {Mazza},
  \citenamefont {Karamatskou}, \citenamefont {Ilchen}, \citenamefont
  {Bakhtiarzadeh}, \citenamefont {Rafipoor}, \citenamefont {O'Keeffe},
  \citenamefont {Kelly}, \citenamefont {Walsh}, \citenamefont {Costello},
  \citenamefont {Meyer},\ and\ \citenamefont {Santra}}]{t.mazza15a}%
  \BibitemOpen
  \bibfield  {author} {\bibinfo {author} {\bibfnamefont {T.}~\bibnamefont
  {Mazza}}, \bibinfo {author} {\bibfnamefont {A.}~\bibnamefont {Karamatskou}},
  \bibinfo {author} {\bibfnamefont {M.}~\bibnamefont {Ilchen}}, \bibinfo
  {author} {\bibfnamefont {S.}~\bibnamefont {Bakhtiarzadeh}}, \bibinfo {author}
  {\bibfnamefont {A.~J.}\ \bibnamefont {Rafipoor}}, \bibinfo {author}
  {\bibfnamefont {P.}~\bibnamefont {O'Keeffe}}, \bibinfo {author}
  {\bibfnamefont {T.~J.}\ \bibnamefont {Kelly}}, \bibinfo {author}
  {\bibfnamefont {N.}~\bibnamefont {Walsh}}, \bibinfo {author} {\bibfnamefont
  {J.~T.}\ \bibnamefont {Costello}}, \bibinfo {author} {\bibfnamefont
  {M.}~\bibnamefont {Meyer}}, \ and\ \bibinfo {author} {\bibfnamefont
  {R.}~\bibnamefont {Santra}},\ }\href {\doibase 10.1038/ncomms7799} {\bibfield
   {journal} {\bibinfo  {journal} {Nat. Commun.}\ }\textbf {\bibinfo {volume}
  {6}},\ \bibinfo {pages} {6799} (\bibinfo {year} {2015})}\BibitemShut
  {NoStop}%
\bibitem [{\citenamefont {Chen}\ \emph {et~al.}(2015)\citenamefont {Chen},
  \citenamefont {Pabst}, \citenamefont {Karamatskou},\ and\ \citenamefont
  {Santra}}]{y.chen15a}%
  \BibitemOpen
  \bibfield  {author} {\bibinfo {author} {\bibfnamefont {Y.-J.}\ \bibnamefont
  {Chen}}, \bibinfo {author} {\bibfnamefont {S.}~\bibnamefont {Pabst}},
  \bibinfo {author} {\bibfnamefont {A.}~\bibnamefont {Karamatskou}}, \ and\
  \bibinfo {author} {\bibfnamefont {R.}~\bibnamefont {Santra}},\ }\href
  {\doibase 10.1103/PhysRevA.91.032503} {\bibfield  {journal} {\bibinfo
  {journal} {Phys. Rev. A}\ }\textbf {\bibinfo {volume} {91}},\ \bibinfo
  {pages} {032503} (\bibinfo {year} {2015})}\BibitemShut {NoStop}%
\bibitem [{\citenamefont {Moiseyev}(2011)}]{n.moiseyev11a}%
  \BibitemOpen
  \bibfield  {author} {\bibinfo {author} {\bibfnamefont {N.}~\bibnamefont
  {Moiseyev}},\ }\href@noop {} {\emph {\bibinfo {title} {{Non-Hermitian Quantum
  Mechanics}}}}\ (\bibinfo  {publisher} {Cambridge University Press},\ \bibinfo
  {address} {New York},\ \bibinfo {year} {2011})\BibitemShut {NoStop}%
\bibitem [{\citenamefont {Moiseyev}(1998{\natexlab{a}})}]{n.moiseyev98b}%
  \BibitemOpen
  \bibfield  {author} {\bibinfo {author} {\bibfnamefont {N.}~\bibnamefont
  {Moiseyev}},\ }\href@noop {} {\bibfield  {journal} {\bibinfo  {journal}
  {Phys. Rep.}\ }\textbf {\bibinfo {volume} {302}},\ \bibinfo {pages} {212}
  (\bibinfo {year} {1998}{\natexlab{a}})}\BibitemShut {NoStop}%
\bibitem [{\citenamefont {Ho}(1983)}]{y.ho83a}%
  \BibitemOpen
  \bibfield  {author} {\bibinfo {author} {\bibfnamefont {Y.~K.}\ \bibnamefont
  {Ho}},\ }\href {\doibase 10.1016/0370-1573(83)90112-6} {\bibfield  {journal}
  {\bibinfo  {journal} {Phys. Rep.}\ }\textbf {\bibinfo {volume} {99}},\
  \bibinfo {pages} {1} (\bibinfo {year} {1983})}\BibitemShut {NoStop}%
\bibitem [{\citenamefont {Reinhardt}(1982)}]{w.reinhardt82a}%
  \BibitemOpen
  \bibfield  {author} {\bibinfo {author} {\bibfnamefont {W.~P.}\ \bibnamefont
  {Reinhardt}},\ }\href@noop {} {\bibfield  {journal} {\bibinfo  {journal}
  {Annu. Rev. Phys. Chem.}\ }\textbf {\bibinfo {volume} {33}},\ \bibinfo
  {pages} {223} (\bibinfo {year} {1982})}\BibitemShut {NoStop}%
\bibitem [{\citenamefont {Santra}\ and\ \citenamefont
  {Cederbaum}(2002)}]{r.santra02a}%
  \BibitemOpen
  \bibfield  {author} {\bibinfo {author} {\bibfnamefont {R.}~\bibnamefont
  {Santra}}\ and\ \bibinfo {author} {\bibfnamefont {L.~S.}\ \bibnamefont
  {Cederbaum}},\ }\href {\doibase 10.1016/S0370-1573(02)00143-6} {\bibfield
  {journal} {\bibinfo  {journal} {Phys. Rep.}\ }\textbf {\bibinfo {volume}
  {368}},\ \bibinfo {pages} {1} (\bibinfo {year} {2002})}\BibitemShut {NoStop}%
\bibitem [{\citenamefont {Altun}\ \emph {et~al.}(1988)\citenamefont {Altun},
  \citenamefont {Kutzner},\ and\ \citenamefont {Kelly}}]{z.altun88a}%
  \BibitemOpen
  \bibfield  {author} {\bibinfo {author} {\bibfnamefont {Z.}~\bibnamefont
  {Altun}}, \bibinfo {author} {\bibfnamefont {M.}~\bibnamefont {Kutzner}}, \
  and\ \bibinfo {author} {\bibfnamefont {H.~P.}\ \bibnamefont {Kelly}},\
  }\href@noop {} {\bibfield  {journal} {\bibinfo  {journal} {Phys. Rev. A}\
  }\textbf {\bibinfo {volume} {37}},\ \bibinfo {pages} {4671} (\bibinfo {year}
  {1988})}\BibitemShut {NoStop}%
\bibitem [{\citenamefont {Cherepkov}\ and\ \citenamefont
  {Semenov}(2001)}]{n.cherepkov01a}%
  \BibitemOpen
  \bibfield  {author} {\bibinfo {author} {\bibfnamefont {N.~A.}\ \bibnamefont
  {Cherepkov}}\ and\ \bibinfo {author} {\bibfnamefont {S.~K.}\ \bibnamefont
  {Semenov}},\ }\href {\doibase 10.1088/0953-4075/34/15/105} {\bibfield
  {journal} {\bibinfo  {journal} {J. Phys. B At. Mol. Phys.}\ }\textbf
  {\bibinfo {volume} {34}},\ \bibinfo {pages} {L495} (\bibinfo {year}
  {2001})}\BibitemShut {NoStop}%
\bibitem [{\citenamefont {Pi}\ and\ \citenamefont {Starace}(2014)}]{l.pi14a}%
  \BibitemOpen
  \bibfield  {author} {\bibinfo {author} {\bibfnamefont {L.-W.}\ \bibnamefont
  {Pi}}\ and\ \bibinfo {author} {\bibfnamefont {A.~F.}\ \bibnamefont
  {Starace}},\ }\href {\doibase 10.1103/PhysRevA.90.023403} {\bibfield
  {journal} {\bibinfo  {journal} {Phys. Rev. A}\ }\textbf {\bibinfo {volume}
  {90}},\ \bibinfo {pages} {023403} (\bibinfo {year} {2014})}\BibitemShut
  {NoStop}%
\bibitem [{\citenamefont {Moiseyev}(1998{\natexlab{b}})}]{n.moiseyev98a}%
  \BibitemOpen
  \bibfield  {author} {\bibinfo {author} {\bibfnamefont {N.}~\bibnamefont
  {Moiseyev}},\ }\href@noop {} {\bibfield  {journal} {\bibinfo  {journal} {J.
  Phys. B At. Mol. Opt. Phys.}\ }\textbf {\bibinfo {volume} {31}},\ \bibinfo
  {pages} {1431} (\bibinfo {year} {1998}{\natexlab{b}})}\BibitemShut {NoStop}%
\bibitem [{\citenamefont {Karlsson}(1998)}]{h.karlsson98a}%
  \BibitemOpen
  \bibfield  {author} {\bibinfo {author} {\bibfnamefont {H.~O.}\ \bibnamefont
  {Karlsson}},\ }\href {\doibase 10.1063/1.477598} {\bibfield  {journal}
  {\bibinfo  {journal} {J. Chem. Phys.}\ }\textbf {\bibinfo {volume} {109}},\
  \bibinfo {pages} {9366} (\bibinfo {year} {1998})}\BibitemShut {NoStop}%
\bibitem [{\citenamefont {Buth}\ and\ \citenamefont
  {Santra}(2007)}]{c.buth07b}%
  \BibitemOpen
  \bibfield  {author} {\bibinfo {author} {\bibfnamefont {C.}~\bibnamefont
  {Buth}}\ and\ \bibinfo {author} {\bibfnamefont {R.}~\bibnamefont {Santra}},\
  }\href {\doibase 10.1103/PhysRevA.75.033412} {\bibfield  {journal} {\bibinfo
  {journal} {Phys. Rev. A}\ }\textbf {\bibinfo {volume} {75}},\ \bibinfo
  {pages} {033412} (\bibinfo {year} {2007})}\BibitemShut {NoStop}%
\bibitem [{\citenamefont {Pabst}\ \emph {et~al.}(2016)\citenamefont {Pabst},
  \citenamefont {Sytcheva}, \citenamefont {Geffert},\ and\ \citenamefont
  {Santra}}]{s.pabst16b}%
  \BibitemOpen
  \bibfield  {author} {\bibinfo {author} {\bibfnamefont {S.}~\bibnamefont
  {Pabst}}, \bibinfo {author} {\bibfnamefont {A.}~\bibnamefont {Sytcheva}},
  \bibinfo {author} {\bibfnamefont {O.}~\bibnamefont {Geffert}}, \ and\
  \bibinfo {author} {\bibfnamefont {R.}~\bibnamefont {Santra}},\ }\href
  {\doibase 10.1103/PhysRevA.94.033421} {\bibfield  {journal} {\bibinfo
  {journal} {Phys. Rev. A}\ }\textbf {\bibinfo {volume} {94}},\ \bibinfo
  {pages} {033421} (\bibinfo {year} {2016})}\BibitemShut {NoStop}%
\bibitem [{\citenamefont {Rohringer}\ \emph {et~al.}(2006)\citenamefont
  {Rohringer}, \citenamefont {Gordon},\ and\ \citenamefont
  {Santra}}]{n.rohringer06a}%
  \BibitemOpen
  \bibfield  {author} {\bibinfo {author} {\bibfnamefont {N.}~\bibnamefont
  {Rohringer}}, \bibinfo {author} {\bibfnamefont {A.}~\bibnamefont {Gordon}}, \
  and\ \bibinfo {author} {\bibfnamefont {R.}~\bibnamefont {Santra}},\ }\href
  {\doibase 10.1103/PhysRevA.74.043420} {\bibfield  {journal} {\bibinfo
  {journal} {Phys. Rev. A}\ }\textbf {\bibinfo {volume} {74}},\ \bibinfo
  {pages} {043420} (\bibinfo {year} {2006})}\BibitemShut {NoStop}%
\bibitem [{\citenamefont {Greenman}\ \emph {et~al.}(2010)\citenamefont
  {Greenman}, \citenamefont {Ho}, \citenamefont {Pabst}, \citenamefont
  {Kamarchik}, \citenamefont {Mazziotti},\ and\ \citenamefont
  {Santra}}]{l.greenman10a}%
  \BibitemOpen
  \bibfield  {author} {\bibinfo {author} {\bibfnamefont {L.}~\bibnamefont
  {Greenman}}, \bibinfo {author} {\bibfnamefont {P.~J.}\ \bibnamefont {Ho}},
  \bibinfo {author} {\bibfnamefont {S.}~\bibnamefont {Pabst}}, \bibinfo
  {author} {\bibfnamefont {E.}~\bibnamefont {Kamarchik}}, \bibinfo {author}
  {\bibfnamefont {D.~A.}\ \bibnamefont {Mazziotti}}, \ and\ \bibinfo {author}
  {\bibfnamefont {R.}~\bibnamefont {Santra}},\ }\href {\doibase
  10.1103/PhysRevA.82.023406} {\bibfield  {journal} {\bibinfo  {journal} {Phys.
  Rev. A}\ }\textbf {\bibinfo {volume} {82}},\ \bibinfo {pages} {023406}
  (\bibinfo {year} {2010})}\BibitemShut {NoStop}%
\bibitem [{\citenamefont {Pabst}\ \emph {et~al.}(2012)\citenamefont {Pabst},
  \citenamefont {Greenman}, \citenamefont {Mazziotti},\ and\ \citenamefont
  {Santra}}]{s.pabst12b}%
  \BibitemOpen
  \bibfield  {author} {\bibinfo {author} {\bibfnamefont {S.}~\bibnamefont
  {Pabst}}, \bibinfo {author} {\bibfnamefont {L.}~\bibnamefont {Greenman}},
  \bibinfo {author} {\bibfnamefont {D.~A.}\ \bibnamefont {Mazziotti}}, \ and\
  \bibinfo {author} {\bibfnamefont {R.}~\bibnamefont {Santra}},\ }\href
  {\doibase 10.1103/PhysRevA.85.023411} {\bibfield  {journal} {\bibinfo
  {journal} {Phys. Rev. A}\ }\textbf {\bibinfo {volume} {85}},\ \bibinfo
  {pages} {023411} (\bibinfo {year} {2012})}\BibitemShut {NoStop}%
\bibitem [{\citenamefont {Pabst}(2013)}]{s.pabst13a}%
  \BibitemOpen
  \bibfield  {author} {\bibinfo {author} {\bibfnamefont {S.}~\bibnamefont
  {Pabst}},\ }\href {\doibase 10.1140/epjst/e2013-01819-x} {\bibfield
  {journal} {\bibinfo  {journal} {Eur. Phys. J. Spec. Top.}\ }\textbf {\bibinfo
  {volume} {221}},\ \bibinfo {pages} {1} (\bibinfo {year} {2013})}\BibitemShut
  {NoStop}%
\bibitem [{\citenamefont {Pabst}\ and\ \citenamefont
  {Santra}(2017)}]{s.pabst17a}%
  \BibitemOpen
  \bibfield  {author} {\bibinfo {author} {\bibfnamefont {S.}~\bibnamefont
  {Pabst}}\ and\ \bibinfo {author} {\bibfnamefont {R.}~\bibnamefont {Santra}},\
  }in\ \href {\doibase 10.1515/9783110417265-006} {\emph {\bibinfo {booktitle}
  {Computational Strong-Field Quantum Dynamics}}},\ Vol.~\bibinfo {volume} {5},\ \bibinfo
  {editor} {edited by\ \bibinfo {editor} {\bibfnamefont {D.}~\bibnamefont
  {Bauer}}}\ (\bibinfo  {publisher} {De Gruyter},\ \bibinfo {address}
  {Berlin},\ \bibinfo {year} {2017})\ p.\ \bibinfo {pages} {169}\BibitemShut
  {NoStop}%
\bibitem [{\citenamefont {Szabo}\ and\ \citenamefont
  {Ostlund}(1996)}]{a.szabo96a}%
  \BibitemOpen
  \bibfield  {author} {\bibinfo {author} {\bibfnamefont {A.}~\bibnamefont
  {Szabo}}\ and\ \bibinfo {author} {\bibfnamefont {N.~S.}\ \bibnamefont
  {Ostlund}},\ }\href@noop {} {\emph {\bibinfo {title} {{Modern Quantum
  Chemistry: Introduction to Advanced Electronic Structure Theory}}}}\
  (\bibinfo  {publisher} {Dover},\ \bibinfo {address} {Mineola, New York},\
  \bibinfo {year} {1996})\BibitemShut {NoStop}%
\bibitem [{\citenamefont {Faccial{\`{a}}}\ \emph {et~al.}(2016)\citenamefont
  {Faccial{\`{a}}}, \citenamefont {Pabst}, \citenamefont {Bruner},
  \citenamefont {Ciriolo}, \citenamefont {{De Silvestri}}, \citenamefont
  {Devetta}, \citenamefont {Negro}, \citenamefont {Soifer}, \citenamefont
  {Stagira}, \citenamefont {Dudovich},\ and\ \citenamefont
  {Vozzi}}]{d.facciala16a}%
  \BibitemOpen
  \bibfield  {author} {\bibinfo {author} {\bibfnamefont {D.}~\bibnamefont
  {Faccial{\`{a}}}}, \bibinfo {author} {\bibfnamefont {S.}~\bibnamefont
  {Pabst}}, \bibinfo {author} {\bibfnamefont {B.~D.}\ \bibnamefont {Bruner}},
  \bibinfo {author} {\bibfnamefont {A.~G.}\ \bibnamefont {Ciriolo}}, \bibinfo
  {author} {\bibfnamefont {S.}~\bibnamefont {{De Silvestri}}}, \bibinfo
  {author} {\bibfnamefont {M.}~\bibnamefont {Devetta}}, \bibinfo {author}
  {\bibfnamefont {M.}~\bibnamefont {Negro}}, \bibinfo {author} {\bibfnamefont
  {H.}~\bibnamefont {Soifer}}, \bibinfo {author} {\bibfnamefont
  {S.}~\bibnamefont {Stagira}}, \bibinfo {author} {\bibfnamefont
  {N.}~\bibnamefont {Dudovich}}, \ and\ \bibinfo {author} {\bibfnamefont
  {C.}~\bibnamefont {Vozzi}},\ }\href {\doibase 10.1103/PhysRevLett.117.093902}
  {\bibfield  {journal} {\bibinfo  {journal} {Phys. Rev. Lett.}\ }\textbf
  {\bibinfo {volume} {117}},\ \bibinfo {pages} {093902} (\bibinfo {year}
  {2016})}\BibitemShut {NoStop}%
\bibitem [{\citenamefont {Krebs}\ \emph {et~al.}(2014)\citenamefont {Krebs},
  \citenamefont {Pabst},\ and\ \citenamefont {Santra}}]{d.krebs14a}%
  \BibitemOpen
  \bibfield  {author} {\bibinfo {author} {\bibfnamefont {D.}~\bibnamefont
  {Krebs}}, \bibinfo {author} {\bibfnamefont {S.}~\bibnamefont {Pabst}}, \ and\
  \bibinfo {author} {\bibfnamefont {R.}~\bibnamefont {Santra}},\ }\href
  {\doibase 10.1119/1.4827015} {\bibfield  {journal} {\bibinfo  {journal} {Am.
  J. Phys.}\ }\textbf {\bibinfo {volume} {82}},\ \bibinfo {pages} {113}
  (\bibinfo {year} {2014})}\BibitemShut {NoStop}%
\bibitem [{\citenamefont {Pabst}\ \emph {et~al.}(2015)\citenamefont {Pabst},
  \citenamefont {Wang},\ and\ \citenamefont {Santra}}]{s.pabst15a}%
  \BibitemOpen
  \bibfield  {author} {\bibinfo {author} {\bibfnamefont {S.}~\bibnamefont
  {Pabst}}, \bibinfo {author} {\bibfnamefont {D.}~\bibnamefont {Wang}}, \ and\
  \bibinfo {author} {\bibfnamefont {R.}~\bibnamefont {Santra}},\ }\href
  {\doibase 10.1103/PhysRevA.92.053424} {\bibfield  {journal} {\bibinfo
  {journal} {Phys. Rev. A}\ }\textbf {\bibinfo {volume} {92}},\ \bibinfo
  {pages} {053424} (\bibinfo {year} {2015})}\BibitemShut {NoStop}%
\bibitem [{\citenamefont {Scrinzi}(2000)}]{a.scrinzi00a}%
  \BibitemOpen
  \bibfield  {author} {\bibinfo {author} {\bibfnamefont {A.}~\bibnamefont
  {Scrinzi}},\ }\href {\doibase 10.1103/PhysRevA.61.041402} {\bibfield
  {journal} {\bibinfo  {journal} {Phys. Rev. A}\ }\textbf {\bibinfo {volume}
  {61}},\ \bibinfo {pages} {041402(R)} (\bibinfo {year} {2000})}\BibitemShut
  {NoStop}%
\bibitem [{\citenamefont {Telnov}\ and\ \citenamefont
  {Chu}(2002)}]{d.telnov02a}%
  \BibitemOpen
  \bibfield  {author} {\bibinfo {author} {\bibfnamefont {D.~A.}\ \bibnamefont
  {Telnov}}\ and\ \bibinfo {author} {\bibfnamefont {S.-I.}\ \bibnamefont
  {Chu}},\ }\href {\doibase 10.1103/PhysRevA.66.043417} {\bibfield  {journal}
  {\bibinfo  {journal} {Phys. Rev. A}\ }\textbf {\bibinfo {volume} {66}},\
  \bibinfo {pages} {043417} (\bibinfo {year} {2002})}\BibitemShut {NoStop}%
\bibitem [{\citenamefont {Bian}\ and\ \citenamefont
  {Bandrauk}(2011)}]{x.bian11a}%
  \BibitemOpen
  \bibfield  {author} {\bibinfo {author} {\bibfnamefont {X.-B.}\ \bibnamefont
  {Bian}}\ and\ \bibinfo {author} {\bibfnamefont {A.~D.}\ \bibnamefont
  {Bandrauk}},\ }\href {\doibase 10.1103/PhysRevA.83.023414} {\bibfield
  {journal} {\bibinfo  {journal} {Phys. Rev. A}\ }\textbf {\bibinfo {volume}
  {83}},\ \bibinfo {pages} {023414} (\bibinfo {year} {2011})}\BibitemShut
  {NoStop}%
\bibitem [{\citenamefont {Pabst}\ \emph {et~al.}(2014)\citenamefont {Pabst},
  \citenamefont {Greenman}, \citenamefont {Karamatskou}, \citenamefont {Chen},
  \citenamefont {Sytcheva}, \citenamefont {Geffert},\ and\ \citenamefont
  {Santra}}]{xcid}%
  \BibitemOpen
  \bibfield  {author} {\bibinfo {author} {\bibfnamefont {S.}~\bibnamefont
  {Pabst}}, \bibinfo {author} {\bibfnamefont {L.}~\bibnamefont {Greenman}},
  \bibinfo {author} {\bibfnamefont {A.}~\bibnamefont {Karamatskou}}, \bibinfo
  {author} {\bibfnamefont {Y.-J.}\ \bibnamefont {Chen}}, \bibinfo {author}
  {\bibfnamefont {A.}~\bibnamefont {Sytcheva}}, \bibinfo {author}
  {\bibfnamefont {O.}~\bibnamefont {Geffert}}, \ and\ \bibinfo {author}
  {\bibfnamefont {R.}~\bibnamefont {Santra}},\ }\href@noop {} {\emph {\bibinfo
  {title} {{\textsc{XCID}---The Configuration-Interaction Dynamics
  Package}}}},\ \bibinfo {edition} {Rev.~1220}\ (\bibinfo  {publisher} {CFEL,
  DESY},\ \bibinfo {address} {Hamburg},\ \bibinfo {year} {2014})\BibitemShut
  {NoStop}%
\bibitem [{\citenamefont {Thompson}(2009)}]{a.thompson09a}%
  \BibitemOpen
  \bibinfo {editor} {\bibfnamefont {A.~C.}\ \bibnamefont {Thompson}},\ ed.,\
  \href {http://xdb.lbl.gov} {\emph {\bibinfo {title} {{X-Ray Data
  Booklet}}}},\ \bibinfo {edition} {3rd}\ ed.\ (\bibinfo  {publisher} {Lawrence
  Berkeley National Laboratory},\ \bibinfo {year} {2009})\BibitemShut {NoStop}%
\bibitem [{\citenamefont {Faisal}(1987)}]{f.faisal87a}%
  \BibitemOpen
  \bibfield  {author} {\bibinfo {author} {\bibfnamefont {F.~H.~M.}\
  \bibnamefont {Faisal}},\ }\href@noop {} {\emph {\bibinfo {title} {{Theory of
  Multiphoton Processes}}}}\ (\bibinfo  {publisher} {Plenum},\ \bibinfo
  {address} {New York},\ \bibinfo {year} {1987})\BibitemShut {NoStop}%
\bibitem [{\citenamefont {Povh}\ \emph {et~al.}(1995)\citenamefont {Povh},
  \citenamefont {Rith}, \citenamefont {Scholz}, \citenamefont {Zetsche},\ and\
  \citenamefont {Rodejohann}}]{b.povh95a}%
  \BibitemOpen
  \bibfield  {author} {\bibinfo {author} {\bibfnamefont {B.}~\bibnamefont
  {Povh}}, \bibinfo {author} {\bibfnamefont {K.}~\bibnamefont {Rith}}, \bibinfo
  {author} {\bibfnamefont {C.}~\bibnamefont {Scholz}}, \bibinfo {author}
  {\bibfnamefont {F.}~\bibnamefont {Zetsche}}, \ and\ \bibinfo {author}
  {\bibfnamefont {W.}~\bibnamefont {Rodejohann}},\ }\href@noop {} {\emph
  {\bibinfo {title} {{Particles and Nuclei: An Introduction to the Physical
  Concepts}}}}\ (\bibinfo  {publisher} {Springer-Verlag},\ \bibinfo {address}
  {Berlin},\ \bibinfo {year} {1995})\BibitemShut {NoStop}%
\bibitem [{\citenamefont {Feldhaus}(2010)}]{j.feldhaus10a}%
  \BibitemOpen
  \bibfield  {author} {\bibinfo {author} {\bibfnamefont {J.}~\bibnamefont
  {Feldhaus}},\ }\href {\doibase 10.1088/0953-4075/43/19/194002} {\bibfield
  {journal} {\bibinfo  {journal} {J. Phys. B At. Mol. Opt. Phys.}\ }\textbf
  {\bibinfo {volume} {43}},\ \bibinfo {pages} {194002} (\bibinfo {year}
  {2010})}\BibitemShut {NoStop}%
\bibitem [{\citenamefont {Allaria}\ \emph {et~al.}(2010)\citenamefont
  {Allaria}, \citenamefont {Callegari}, \citenamefont {Cocco}, \citenamefont
  {Fawley}, \citenamefont {Kiskinova}, \citenamefont {Masciovecchio},\ and\
  \citenamefont {Parmigiani}}]{e.allaria10a}%
  \BibitemOpen
  \bibfield  {author} {\bibinfo {author} {\bibfnamefont {E.}~\bibnamefont
  {Allaria}}, \bibinfo {author} {\bibfnamefont {C.}~\bibnamefont {Callegari}},
  \bibinfo {author} {\bibfnamefont {D.}~\bibnamefont {Cocco}}, \bibinfo
  {author} {\bibfnamefont {W.~M.}\ \bibnamefont {Fawley}}, \bibinfo {author}
  {\bibfnamefont {M.}~\bibnamefont {Kiskinova}}, \bibinfo {author}
  {\bibfnamefont {C.}~\bibnamefont {Masciovecchio}}, \ and\ \bibinfo {author}
  {\bibfnamefont {F.}~\bibnamefont {Parmigiani}},\ }\href {\doibase
  10.1088/1367-2630/12/7/075002} {\bibfield  {journal} {\bibinfo  {journal}
  {New J. Phys.}\ }\textbf {\bibinfo {volume} {12}},\ \bibinfo {pages} {075002}
  (\bibinfo {year} {2010})}\BibitemShut {NoStop}%
\bibitem [{\citenamefont {Sekikawa}\ \emph {et~al.}(2004)\citenamefont
  {Sekikawa}, \citenamefont {Kosuge}, \citenamefont {Kanai},\ and\
  \citenamefont {Watanabe}}]{t.sekikawa04a}%
  \BibitemOpen
  \bibfield  {author} {\bibinfo {author} {\bibfnamefont {T.}~\bibnamefont
  {Sekikawa}}, \bibinfo {author} {\bibfnamefont {A.}~\bibnamefont {Kosuge}},
  \bibinfo {author} {\bibfnamefont {T.}~\bibnamefont {Kanai}}, \ and\ \bibinfo
  {author} {\bibfnamefont {S.}~\bibnamefont {Watanabe}},\ }\href {\doibase
  10.1038/nature03074.1.} {\bibfield  {journal} {\bibinfo  {journal} {Nature}\ }\textbf {\bibinfo {volume} {432}},\ \bibinfo {pages} {605}
  (\bibinfo {year} {2004})}\BibitemShut {NoStop}%
\bibitem [{\citenamefont {Krausz}\ and\ \citenamefont
  {Ivanov}(2009)}]{f.krausz09a}%
  \BibitemOpen
  \bibfield  {author} {\bibinfo {author} {\bibfnamefont {F.}~\bibnamefont
  {Krausz}}\ and\ \bibinfo {author} {\bibfnamefont {M.}~\bibnamefont
  {Ivanov}},\ }\href {\doibase 10.1103/RevModPhys.81.163} {\bibfield  {journal}
  {\bibinfo  {journal} {Rev. Mod. Phys.}\ }\textbf {\bibinfo {volume} {81}},\
  \bibinfo {pages} {163} (\bibinfo {year} {2009})}\BibitemShut {NoStop}%
\bibitem [{\citenamefont {Rescigno}\ and\ \citenamefont
  {McKoy}(1975)}]{t.rescigno75a}%
  \BibitemOpen
  \bibfield  {author} {\bibinfo {author} {\bibfnamefont {T.~N.}\ \bibnamefont
  {Rescigno}}\ and\ \bibinfo {author} {\bibfnamefont {V.}~\bibnamefont
  {McKoy}},\ }\href {\doibase 10.1103/PhysRevA.12.522} {\bibfield  {journal}
  {\bibinfo  {journal} {Phys. Rev. A}\ }\textbf {\bibinfo {volume} {12}},\
  \bibinfo {pages} {522} (\bibinfo {year} {1975})}\BibitemShut {NoStop}%
\bibitem [{\citenamefont {Jurvansuu}\ \emph {et~al.}(2001)\citenamefont
  {Jurvansuu}, \citenamefont {Kivim{\"{a}}ki},\ and\ \citenamefont
  {Aksela}}]{m.jurvansuu01a}%
  \BibitemOpen
  \bibfield  {author} {\bibinfo {author} {\bibfnamefont {M.}~\bibnamefont
  {Jurvansuu}}, \bibinfo {author} {\bibfnamefont {A.}~\bibnamefont
  {Kivim{\"{a}}ki}}, \ and\ \bibinfo {author} {\bibfnamefont {S.}~\bibnamefont
  {Aksela}},\ }\href {\doibase 10.1103/PhysRevA.64.012502} {\bibfield
  {journal} {\bibinfo  {journal} {Phys. Rev. A}\ }\textbf {\bibinfo {volume}
  {64}},\ \bibinfo {pages} {012502} (\bibinfo {year} {2001})}\BibitemShut
  {NoStop}%
\bibitem [{\citenamefont {Huppert}\ \emph {et~al.}(2016)\citenamefont
  {Huppert}, \citenamefont {Jordan}, \citenamefont {Baykusheva}, \citenamefont
  {von Conta},\ and\ \citenamefont {W{\"{o}}rner}}]{m.huppert16a}%
  \BibitemOpen
  \bibfield  {author} {\bibinfo {author} {\bibfnamefont {M.}~\bibnamefont
  {Huppert}}, \bibinfo {author} {\bibfnamefont {I.}~\bibnamefont {Jordan}},
  \bibinfo {author} {\bibfnamefont {D.}~\bibnamefont {Baykusheva}}, \bibinfo
  {author} {\bibfnamefont {A.}~\bibnamefont {von Conta}}, \ and\ \bibinfo
  {author} {\bibfnamefont {H.~J.}\ \bibnamefont {W{\"{o}}rner}},\ }\href
  {\doibase 10.1103/PhysRevLett.117.093001} {\bibfield  {journal} {\bibinfo
  {journal} {Phys. Rev. Lett.}\ }\textbf {\bibinfo {volume} {117}},\ \bibinfo
  {pages} {093001} (\bibinfo {year} {2016})}\BibitemShut {NoStop}%
\bibitem [{\citenamefont {Ossiander}\ \emph {et~al.}(2017)\citenamefont
  {Ossiander}, \citenamefont {Siegrist}, \citenamefont {Shirvanyan},
  \citenamefont {Pazourek}, \citenamefont {Sommer}, \citenamefont {Latka},
  \citenamefont {Guggenmos}, \citenamefont {Nagele}, \citenamefont {Feist},
  \citenamefont {Burgd{\"{o}}rfer}, \citenamefont {Kienberger},\ and\
  \citenamefont {Schultze}}]{m.ossiander17a}%
  \BibitemOpen
  \bibfield  {author} {\bibinfo {author} {\bibfnamefont {M.}~\bibnamefont
  {Ossiander}}, \bibinfo {author} {\bibfnamefont {F.}~\bibnamefont {Siegrist}},
  \bibinfo {author} {\bibfnamefont {V.}~\bibnamefont {Shirvanyan}}, \bibinfo
  {author} {\bibfnamefont {R.}~\bibnamefont {Pazourek}}, \bibinfo {author}
  {\bibfnamefont {A.}~\bibnamefont {Sommer}}, \bibinfo {author} {\bibfnamefont
  {T.}~\bibnamefont {Latka}}, \bibinfo {author} {\bibfnamefont
  {A.}~\bibnamefont {Guggenmos}}, \bibinfo {author} {\bibfnamefont
  {S.}~\bibnamefont {Nagele}}, \bibinfo {author} {\bibfnamefont
  {J.}~\bibnamefont {Feist}}, \bibinfo {author} {\bibfnamefont
  {J.}~\bibnamefont {Burgd{\"{o}}rfer}}, \bibinfo {author} {\bibfnamefont
  {R.}~\bibnamefont {Kienberger}}, \ and\ \bibinfo {author} {\bibfnamefont
  {M.}~\bibnamefont {Schultze}},\ }\href {\doibase 10.1038/nphys3941}
  {\bibfield  {journal} {\bibinfo  {journal} {Nat. Phys.}\ }\textbf {\bibinfo
  {volume} {13}},\ \bibinfo {pages} {280} (\bibinfo {year} {2017})}\BibitemShut {NoStop}%
\bibitem [{\citenamefont {Kramida}\ \emph {et~al.}(2014)\citenamefont
  {Kramida}, \citenamefont {Ralchenko}, \citenamefont {Reader},\ and\
  \citenamefont {{NIST ASD Team}}}]{a.kramida14a}%
  \BibitemOpen
  \bibfield  {author} {\bibinfo {author} {\bibfnamefont {A.}~\bibnamefont
  {Kramida}}, \bibinfo {author} {\bibfnamefont {Y.}~\bibnamefont {Ralchenko}},
  \bibinfo {author} {\bibfnamefont {J.}~\bibnamefont {Reader}}, \ and\ \bibinfo
  {author} {\bibnamefont {{NIST ASD Team}}},\ }\href
  {http://www.nist.gov/pml/data/asd.cfm} {\emph {\bibinfo {title} {{NIST Atomic
  Spectra Database}}}},\ \bibinfo {edition} {5th}\ ed.\ (\bibinfo  {publisher}
  {National Institute of Standards and Technology},\ \bibinfo {address}
  {Gaithersburg, Maryland},\ \bibinfo {year} {2014})\BibitemShut {NoStop}%
\bibitem [{\citenamefont {Riss}\ and\ \citenamefont {Meyer}(1993)}]{u.riss93a}%
  \BibitemOpen
  \bibfield  {author} {\bibinfo {author} {\bibfnamefont {U.~V.}\ \bibnamefont
  {Riss}}\ and\ \bibinfo {author} {\bibfnamefont {H.-D.}\ \bibnamefont
  {Meyer}},\ }\href@noop {} {\bibfield  {journal} {\bibinfo  {journal} {J.
  Phys. B At. Mol. Opt. Phys.}\ }\textbf {\bibinfo {volume} {26}},\ \bibinfo
  {pages} {4503} (\bibinfo {year} {1993})}\BibitemShut {NoStop}%
\bibitem [{\citenamefont {Sorensen}\ \emph {et~al.}(1996)\citenamefont
  {Sorensen}, \citenamefont {Lehoucq}, \citenamefont {Yang},\ and\
  \citenamefont {Maschhoff}}]{d.sorensen96a}%
  \BibitemOpen
  \bibfield  {author} {\bibinfo {author} {\bibfnamefont {D.~C.}\ \bibnamefont
  {Sorensen}}, \bibinfo {author} {\bibfnamefont {R.~B.}\ \bibnamefont
  {Lehoucq}}, \bibinfo {author} {\bibfnamefont {C.}~\bibnamefont {Yang}}, \
  and\ \bibinfo {author} {\bibfnamefont {K.}~\bibnamefont {Maschhoff}},\ }\href
  {http://www.caam.rice.edu/software/ARPACK/} {\emph {\bibinfo {title}
  {{\textsc{ARPACK}--ARnoldi PACKage}}}},\ \bibinfo {edition} {Ver.~2.1}\
  (\bibinfo  {publisher} {Rice University},\ \bibinfo {address} {Houston,
  Texas},\ \bibinfo {year} {1996})\BibitemShut {NoStop}%
\bibitem [{\citenamefont {Krebs}(2013)}]{d.krebs13a}%
  \BibitemOpen
  \bibfield  {author} {\bibinfo {author} {\bibfnamefont {D.}~\bibnamefont
  {Krebs}},\ }\emph {\bibinfo {title} {{Atomic photoabsorption spectroscopy
  using time-dependent configuration-interaction singles}}},\ \href@noop {}
  {\bibinfo {type} {Bachelor thesis}},\ \bibinfo  {school} {University of
  Hamburg} (\bibinfo {year} {2013})\BibitemShut {NoStop}%
\end{thebibliography}%

\end{document}